\documentclass[reprint,amsmath,amssymb,superscriptaddress,prb]{revtex4-2}

\usepackage{graphicx}
\usepackage{dcolumn}
\usepackage{bm}
\usepackage{color}

\begin{document}

\title{Description of the hydrogen atom and the He$^+$ ion in an
optical cavity using the Pauli-Fierz Hamiltonian}

\author{Yetmgeta S Aklilu}
\affiliation{Department of Physics and Astronomy, Vanderbilt
University, Nashville, Tennessee, 37235, USA}

\author{K\'alm\'an Varga}
\email{kalman.varga@vanderbilt.edu}
\affiliation{Department of Physics and Astronomy, Vanderbilt
University, Nashville, Tennessee, 37235, USA}

\begin{abstract}
A system of one electron in a Coulomb potential in an optical cavity
is solved using a tensor-product light-matter basis. The problem was
treated at the level of the Pauli-Fierz Hamiltonian describing both
light and matter quantum mechanically. The effect of cavity
size on the energy levels and high harmonics generation (HHG) spectrum is
studied. We have shown that the energy levels, transition states, 
entanglement, and the HHG spectrum can be strongly influenced by
changing the cavity size.
\end{abstract}
\maketitle

\section{Introduction}
The interaction of atoms or molecules and light is
usually described using a quantum description for the matter and
a classical treatment for the   electromagnetic fields.
The potential realization of new phases of matter by engineering ultrastrong
light-matter coupling
led to the development of tunable ultra-small optical cavities
\cite{Aupiais2023,doi:10.1021/acs.nanolett.2c01695}. Understanding
the physics of these systems requires describing  the interaction
of  matter and light at the quantum level 
\cite{Bhattacharya_2023,Gorlach2023,Bogatskaya2017,PhysRevA.108.063101,
PhysRevA.109.033706,Moiseyev_2024,Bogatskaya_2020,PhysRevA.108.053119,
doi:10.1021/acs.accounts.6b00295,doi:10.1146/annurev-physchem-090519-042621,
Ruggenthaler2018,mctague,doi:10.1021/acsphotonics.2c00048,rd,PhysRevLett.127.273601,
PhysRevLett.128.156402,10.1063/5.0123909,PhysRevB.107.235130,PhysRevB.105.115127,
PhysRevB.104.165147}. The need for quantum description has been
discussed in many research works. In Refs.
\cite{Bogatskaya_2017,Bogatskaya_2020} the spontaneous
emission driven by a high-intensity laser pulse was investigated using
perturbation theory and including the vacuum-quantized field modes.
The importance of the quantum state of light in high harmonic
generation  was emphasized in Ref. \cite{Gorlach2023}. The
quantum optical aspects of HHG were also studied
\cite{PhysRevA.101.013418,photonics8070269,PhysRevA.104.033703,Gorlach2020,PhysRevA.109.033706}.
The studies of the strong field ionization of atoms by quantum light 
\cite{PhysRevLett.130.253201,PhysRevA.108.063101}  predict new interference patterns
for the tunneling electrons. The light-matter entanglement in
ionization processes was also explored \cite{PhysRevA.106.063705}.

Model systems describing the light-matter interactions with simple
analytical or numerical solutions have always played
important roles in understanding complex physical
systems. The interaction of light and
atoms, for example, can be described 
by the Jaynes-Cummings model \cite{Jaynes1962ComparisonOQ}, 
which assumes a two-level atom is weakly coupled to a single mode of a
quantized electromagnetic field. For 
strongly coupled light-matter systems
\cite{doi:10.1021/acsphotonics.9b00648,Schafer4883,Ruggenthaler2018,Flick15285,
Flick3026,Rokaj_2018,PhysRevLett.122.193603,PhysRevLett.121.113002,doi:10.1063/5.0012723,
PhysRevB.98.235123,PhysRevLett.119.136001,Mandal,Cederbaum2021,
doi:10.1021/acs.jpclett.8b02609,PhysRevLett.126.153603} no simple
approach exists and the light-matter coupling cannot be treated perturbatively either. 

In this work, we will study the properties of a one-electron atom or
ion in a cavity using the Pauli-Fierz (PF) Hamiltonian.
The PF Hamiltonian describes the interaction between quantum matters
(electrons) and a massless quantized radiation field (photons) in the
low-energy non-relativistic limit of quantum electrodynamics (QED)
\cite{craig1984molecular}. The PF approach has often been used in
describing the modification of material properties in optical cavities
\cite{Ruggenthaler2018,Rokaj_2018,Mandal,acs.jpcb.0c03227,PhysRevB.98.235123}.

We will use a Gaussian basis to represent the spatial variables and
a Fock basis to represent the photons. The Gaussian basis is flexible enough
to accurately describe the electronic states not only for the ground
state but also for strong laser excitations. The Hamiltonian is
diagonalized on a basis formed by the tensor product of
the Gaussian states and the Fock states. We will study the dependence
of the energy levels and the light-matter interaction on the cavity size.
The cavity size determines the cavity frequency and the coupling
strength between light and matter.

We will use time-dependent excitations to study the high-harmonic
generation (HHG) and the absorption spectrum in a cavity. In this case the 
time-dependent Schr\"odinger equation will be solved by time
propagation and the absorption and HHG spectrum will be calculated
using the time-dependent dipole moment.
HHG is  a coherent source of extreme ultraviolet emission and many
advances in attoscience are based on HHG. We will show the light-matter
coupling leads to many additional peaks in the HHG spectrum.

\section{Formalism}
We will consider a rectangular cavity with perfectly conducting walls
and use the Coulomb gauge $\boldsymbol{\nabla}\cdot \mathbf{A}=0$. The sides of 
the rectangular cavity are $L_x$, $L_y$, and $L_z$. 
The cavity modes are defined by the momentum vectors 
$k_x={n_x\pi \over L_x}$,
$k_y={n_y\pi \over L_y}$,
$k_z={n_z\pi \over L_z}$, where $n_x,n_y,n_z$ are $0,1,2,{\ldots}$.
The cavity frequency is $\omega_{\mathbf{k}}=c|\mathbf{k}|$ and the mode
functions are 
\begin{equation}
\mathbf{S}_{\mathbf{k}\nu}(\mathbf{r})=\sqrt{2^3\over V}\left(
\begin{array}{c}
\epsilon^{(x)}_{\mathbf{k}\nu}\cos(k_xx)\sin(k_yy)\sin(k_zz)\\
\epsilon^{(y)}_{\mathbf{k}\nu}\sin(k_xx)\cos(k_yy)\sin(k_zz)\\
\epsilon^{(z)}_{\mathbf{k}\nu}\sin(k_xx)\sin(k_yy)\cos(k_zz)\\
\end{array}
\right)
\label{node}
\end{equation}
where $V=L_xL_yL_z$, $\nu$ is the polarization index, and
$\boldsymbol{\epsilon}_{\mathbf{k}\nu}$ is the polarization vector.
These mode functions satisfy the perfect conductor boundary conditions 
$\mathbf{n}\times\mathbf{E}_{\perp}=0$, $\mathbf{n}\times\mathbf{A}=0$.

Using this mode functions the quantized vector potential operator and
electromagnetic fields are defined as
\begin{eqnarray}\hat{ \mathbf{A}}(\mathbf{r}, t) & =&\sum_{\mathbf{k}
\nu} \sqrt{\frac{c^{2}}{\varepsilon_{0}}} \hat{q}_{\mathbf{k}
\nu}(t) \mathbf{S}_{\mathbf{k} \nu}(\mathbf{r}) \nonumber \\
\hat{\mathbf{E}}_{\perp}(\mathbf{r}, t) & =&-\frac{1}{c} \partial_{t}
\hat{\mathbf{A}}(\mathbf{r}, t)\nonumber \\ \hat{\mathbf{B}}(\mathbf{r}, t) & =&\frac{1}{c}
\boldsymbol{\nabla} \times \hat{\mathbf{A}}(\mathbf{r}, t) 
,
\end{eqnarray} 
where $\hat{q}_k\doteq \hat{q}_{\mathbf{k}\nu}$ is connected to the
the creation and annihilation operators
\begin{equation}
\begin{array}{c}
\hat{q}_{k}=\left(\frac{\hbar}{2 \omega_{k}}\right)^{1 /
2}\left(\hat{a}_{k}^{\dagger}+\hat{a}_{k}\right) \\
\hat{p}_{k}=i\left(\frac{\hbar \omega_{k}}{2}\right)^{1 /
2}\left(\hat{a}_{k}^{\dagger}-\hat{a}_{k}\right)
\end{array}
\end{equation}
and $\hat{a}_i$ satisfies $ \left[\hat{a}_{i}, \hat{a}_{j}^{\dagger}\right]=\delta_{i j}$.

The Hamiltonian is defined as
\begin{eqnarray}
\hat{H}& =&\frac{1}{2
m}\left(\hat{p}-\frac{e}{c}
\hat{\mathbf{A}}\right)^{2}-{Z\over r}+V_{ext}(\mathbf{r},t)+\hat{H}_{p h}
\nonumber\\
\hat{H}_{p h} & =&\frac{\epsilon_{0}}{2} \int d \mathbf{r}
\hat{\mathbf{E}}_{\perp}(\mathbf{r})^{2}+c^{2}
\hat{\mathbf{B}}(\mathbf{r})^{2} \nonumber \\
& =&\frac{1}{2} \sum_{{k}}
\hat{p}_{{k}}^{2}+\omega_{k}^{2}\hat{q}_{{k}}^{2},
\end{eqnarray}
where $V_{ext}$ is a time-dependent external field, e.g. a classical
laser field.

One can use a unitary transformation to transform the Hamiltonian into
length gauge \cite{Rokaj_2018,Taylor:22}
\begin{equation}
\hat{H}_L=U^{\dagger}\hat{H}U,
\end{equation}
with 
\begin{equation}
U=\exp\left\lbrace-{i\over \hbar} 
\hat{\mathbf{A}}\cdot \mathbf{D}\right\rbrace,
\end{equation}
where $\mathbf{D}=-e\mathbf{r}$ is the dipole moment.

After a straightforward calculation \cite{Rokaj_2018} one has
\begin{equation}
\hat{H}_L=\hat{H}_m+
\sum_{k}{1\over 2}\left[\hat{p}_{k}^2+ 
\omega_{k}^2
\left(\hat{q}_{k}-{\boldsymbol{\lambda}_{k}\over\omega_{k}}\mathbf{D}\right)^2\right],
\label{tdks1}
\end{equation}
where 
\begin{equation}
\hat{H}_m=-{\hbar^2\over 2m }\boldsymbol{\nabla}^2  -{Z\over r}+
V_{ext}(\mathbf{r},t)
\end{equation}
and
\begin{equation}
\boldsymbol{\lambda}_{k}=
{\mathbf{S}_{\mathbf{k}\nu}(\mathbf{r})\over \sqrt{\epsilon_0}}.
\end{equation}
We will use atomic units in the rest of the paper, $\hbar=m=e=1$,
$\epsilon_0={1\over 4\pi}$ and $c={1\over \alpha}$ (where
$\alpha\approx{1\over 137}$ is
the fine structure constant).

The trial wave function is written as
\begin{equation}
\Phi_i(\mathbf{r})=\sum_{j,\mathbf{n}} a_{i,j}^{\mathbf{n}}
\phi_j^{\mathbf n}(\mathbf{r})
\end{equation}
where 
\begin{equation}
\phi_j^{\mathbf n}(\mathbf{r})=\phi_j(\mathbf{r}) |\mathbf{n}\rangle
\end{equation}
and the Fock state basis
\begin{equation}
|\mathbf{n}\rangle =
|n_1\rangle_{\omega_1}
|n_2\rangle_{\omega_2}{\ldots} 
|n_M\rangle_{\omega_M}.
\label{fock}
\end{equation}

For the calculations we need $H$ the matrix elements of the Hamilton
operator
\begin{equation}
\langle  \phi_i^{\mathbf{n}}\vert H_L \vert \phi_j^{\mathbf{m}}\rangle =
h_{ij}\delta_{\mathbf{n}\mathbf{m}}+c_{ij}^{\mathbf{n}\mathbf{m}}.
\label{hamil}
\end{equation}
The first term is a Hamiltonian where the different Fock states are
not coupled 
\begin{equation}
h_{ij}=\langle  \phi_i\vert 
\hat{H}_m+\sum_{k=1}^M
\left[\omega_k\left(n_k+\frac{1}{2}\right)+
\frac{1}{2}({\boldsymbol{\lambda}}_{k}\cdot\mathbf{D})^2\right]
\vert \phi_j\rangle 
\label{hamilm}
\end{equation}
and the coupling  part
\begin{eqnarray}
c_{ij}^{\mathbf{n}\mathbf{m}}&=&
\langle \phi_i\vert \mathbf{D}\vert \phi_j\rangle 
\sum_{k=1}^M{\boldsymbol{\lambda}}_{k}\omega_k\langle
\mathbf{n}\vert\hat{q}_k\vert\mathbf{m}\rangle\\
&=&
\langle \phi_i\vert \mathbf{D}\vert \phi_j\rangle
\sum_{k=1}^M{\boldsymbol{\lambda}}_{k}{\sqrt{\frac{\omega_k}{2}}}
q_{n_km_k}\prod_{j=1,j\ne k}^M\delta_{n_j,m_j}
\nonumber
\label{coupm}
\end{eqnarray}
with
\begin{equation}
q_{nm}=\sqrt{n}\delta_{nm-1}+\sqrt{n+1}\delta_{nm+1}.
\end{equation}
Finally, the overlap matrix, $O$ of the basis functions are given by
\begin{equation}
\langle  \phi_i^{\mathbf{n}}\vert \phi_j^{\mathbf{m}}\rangle =
\langle  \phi_i\vert \phi_j\rangle 
\delta_{\mathbf{n}\mathbf{m}}.
\label{over}
\end{equation}

If the coupling in Eq. \eqref{coupm} is weak and the self-dipole
interaction (the last term in Eq. \eqref{hamilm}) is small, then the
approximate eigenvalues will be the sum of the eigenenergy of the
Hydrogen atom $E_i$ and the energy of the harmonic oscillators,
$E_i+\sum_{k=1}^M \omega_k(n_k+1/2)$ for each $i$. As we will show
later, $\omega_k$ strongly increases when one decreases the volume of
the cavity. This shifts  the energy levels upwards.
But when the cavity volume becomes smaller the coupling and the 
dipole self-interaction become stronger and this approximation is 
not valid anymore.

Gaussian basis functions 
are used to represent the spatial wave  function:
\begin{equation}
\phi_i(\mathbf{r})=N_ix^{l_i}y^{m_i}z^{n_i}\exp(-\alpha_i \mathbf{r}^2),
\end{equation}
where
\begin{equation}
N_i=\left(\frac{2\alpha}{\pi}\right)^{3/4}\left[
\frac{(8\alpha)^{l+m+n}l! m! n!}{(2l)! (2m)! (2n)!} \right]^{1/2}.
\end{equation}
One could use hydrogenic eigenfunctions for the expansion but the
Gaussians have better flexibility in describing the long density tails
caused by the laser field. The matrix elements of the Gaussian basis
functions are readily available 
\cite{Petersson_2010,doi:10.1098/rspa.1950.0036,RevModPhys.85.693,suzuki1998stochastic}.

For ground-state calculations the generalized eigenvalue problem of
the Hamiltonian and overlap matrix is solved. For time-dependent
calculations the time-dependent Schr\"odinger equation
\begin{equation}
\frac{\partial}{\partial t}\Psi=H_L \Psi
\end{equation}
is solved by time propagation using the Crank-Nicolson method.$\Psi$ 
is defined using the same basis states but with time-dependent
linear combination coefficients
\begin{equation}
\Psi(t)=\sum_{j,\mathbf{n}} a_{j}^{\mathbf{n}}(t)
\phi_j^{\mathbf n}(\mathbf{r}).
\label{tdw}
\end{equation}
The Crank-Nicolson time propagation is defined as
\begin{equation}
C(t+\Delta t)=\frac{O-\frac{1}{2}H\Delta t}{O-\frac{1}{2}H\Delta
t}C(t),
\end{equation}
where $C(t)$ is the vector of the coefficients in Eq. \eqref{tdw} and
the starting wave function $\Psi(t=0)$ is the ground state wave function.

The high harmonic spectrum is calculated using the dipole acceleration:
\begin{equation}
I(\omega)=\left|\int_0^T{\partial^2 d(t)\over \partial
t^2}{\rm e}^{-i\omega t} dt\right|^2, 
\label{hhg}
\end{equation}
\begin{equation}
d(t)=\langle \Psi(t)\vert D_z\vert\Psi(t)\rangle.
\end{equation}
One can also define the component of the dipole moment in a given
photon space by 
\begin{equation}
d_{\mathbf{n}}(t)=\sum_j \left(a_{j}^{\mathbf{n}}(t)\right)^2
\langle \phi_j^{\mathbf n} \vert D_z\vert 
\phi_j^{\mathbf n}\rangle,
\label{pcon}
\end{equation}
and the total dipole will be the sum of these contributions due to
the orthogonality of the Fock space basis functions.
To calculate the high harmonic spectrum, we time propagate a system 
subjected to a classical laser pulse
\begin{equation}
V_{ext}(t)=E_0 {\rm e}^{\frac{(t-t_0)^2}{\tau^2}}\sin(\omega_h t)D_z
\label{laser}
\end{equation}
and  calculate $d(t)$, and then extract $I(\omega)$. 

The time-dependent dipole moment will also be used to calculate the
optical absorption cross-section using
\begin{equation}
S(E)={1\over q}\int (d(t)-d(0)){\rm e}^{iEt} d(t),
\end{equation}
where $q$ is the strength of the exciting  potential
\begin{equation}
V_{ext}(t)=q\delta(t)z.
\label{kick}
\end{equation}

\section{Results and discussion}
We place the atom into the center of the cavity, at
$(L_x/2,L_y/2,L_z/2)$, choose $\boldsymbol{\epsilon}_{k}=(1,0,0)$ 
for the the polarization vector and use $\boldsymbol{\lambda}_k=\lambda \boldsymbol{\epsilon}_{k}$.
To keep $\omega_{{k}}$ close
to the excitation energies of the system one needs large $L_x,L_y$ and
$L_z$, but this leads to weaker coupling (see Eq. \eqref{node}). We
will test rectangular cavities with sides from a few nanometers to a few
hundred nanometers. A recent review of subnanometer gap 
picocavities can be found in Ref. \cite{doi:10.1021/acs.nanolett.2c01695}.

The H atom and the He$^+$ ion will be used in the simulation. 
Eq. \eqref{coupm} shows that nonzero dipole matrix elements couple the
light and the matter states. The simplest example of these is 
the dipole matrix element between the 1s ground and 2p excited state
(there are 3 2p states, we only consider one of them, 2p$_x$ in the
following).
Without the cavity, the energy difference between the 1s and the 2p states 
of the H atom is 0.375 and the energy difference is 1.5 in the case of the He$^+$ ion. 
We want to study the behavior of the system when the cavity
frequency is close to the 1s-2p energy difference. When using $L_x=20,
L_y=L_z=406$ the lowest $\omega_{k}$ is 1.5 and ${\lambda}_k=0.011$.
In order to decrease 
the lowest $\omega_{{k}}$ to 0.375 we will set $L_x=20,L_y=L_z=1623$ 
which gives ${\lambda}_k=0.0028$. As can be seen, decreasing
the frequency makes $\lambda_k$ significantly 
smaller (see Eq. \eqref{node}). The gap size, $L_x=20$, is about 1 nm
and the recently created subnanometer gap \cite{doi:10.1021/acs.nanolett.2c01695}
picocavities make the $L_x=10-20$  choice realistic. Below $L_x=10$
the atomistic details of the cavity wall and the spill of the wave
function of the H atom beyond the cavity wall would make the model
unphysical.

\begin{figure}
\includegraphics[width=0.45\textwidth]{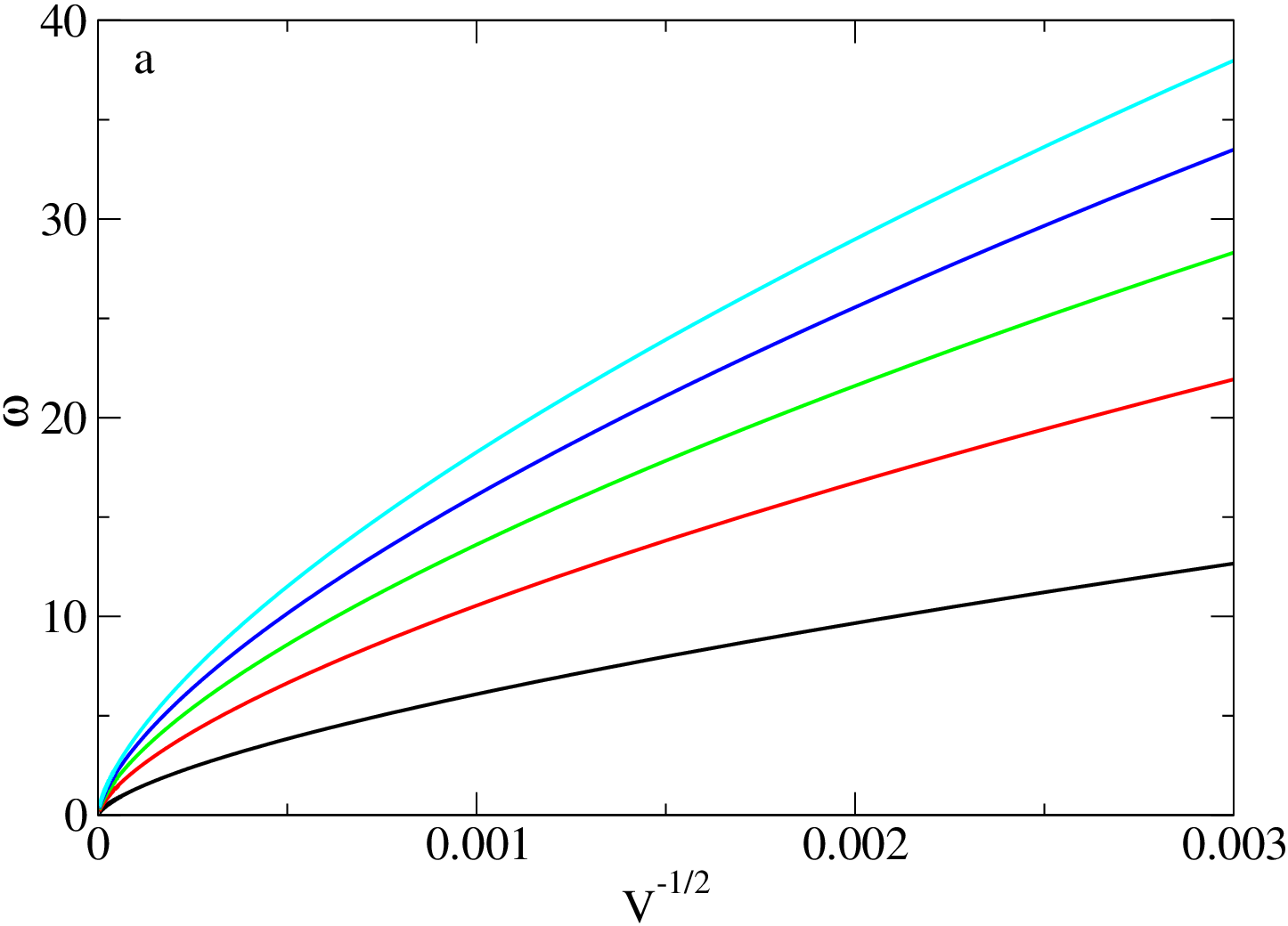}\\
\includegraphics[width=0.45\textwidth]{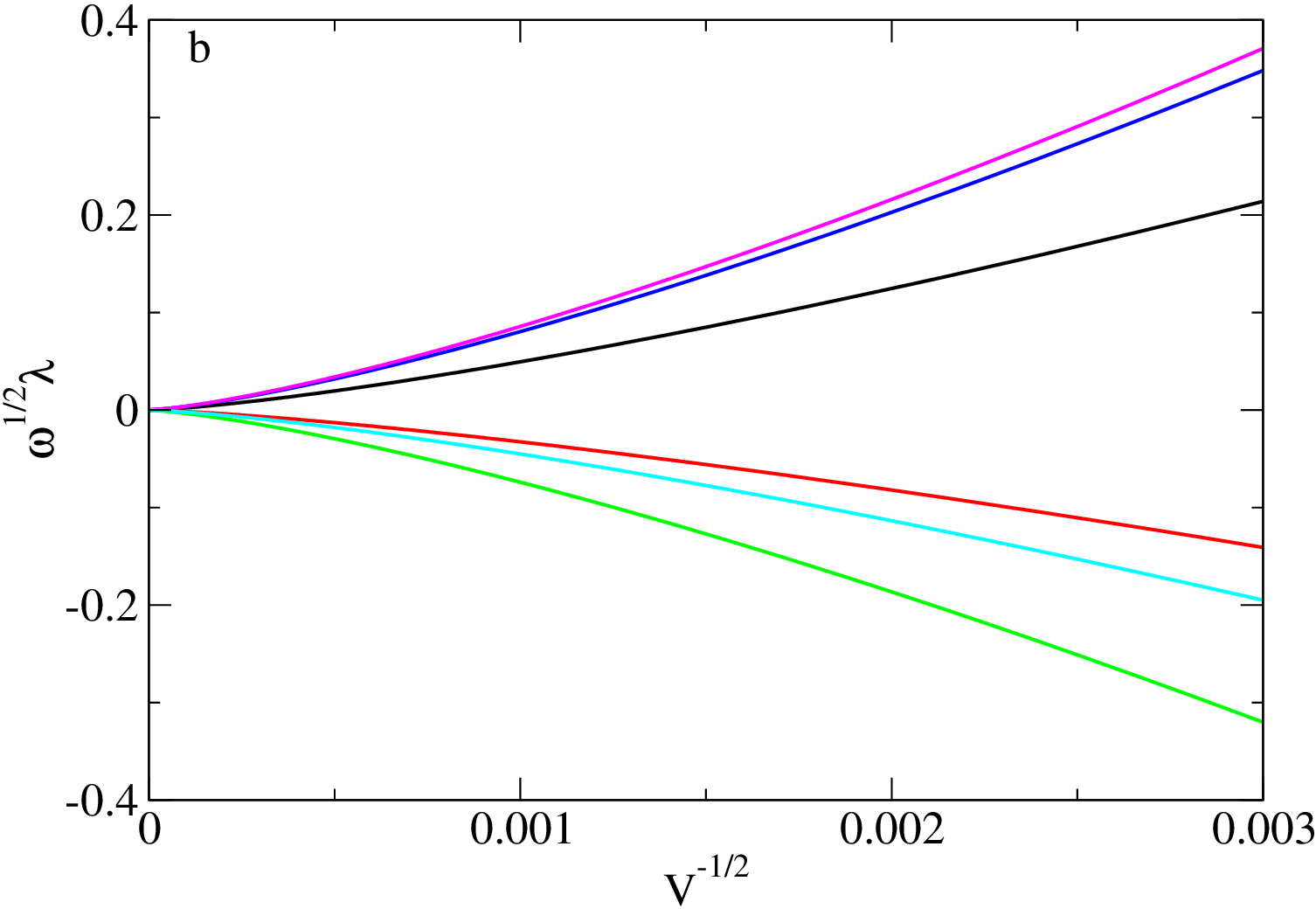}\\
\includegraphics[width=0.45\textwidth]{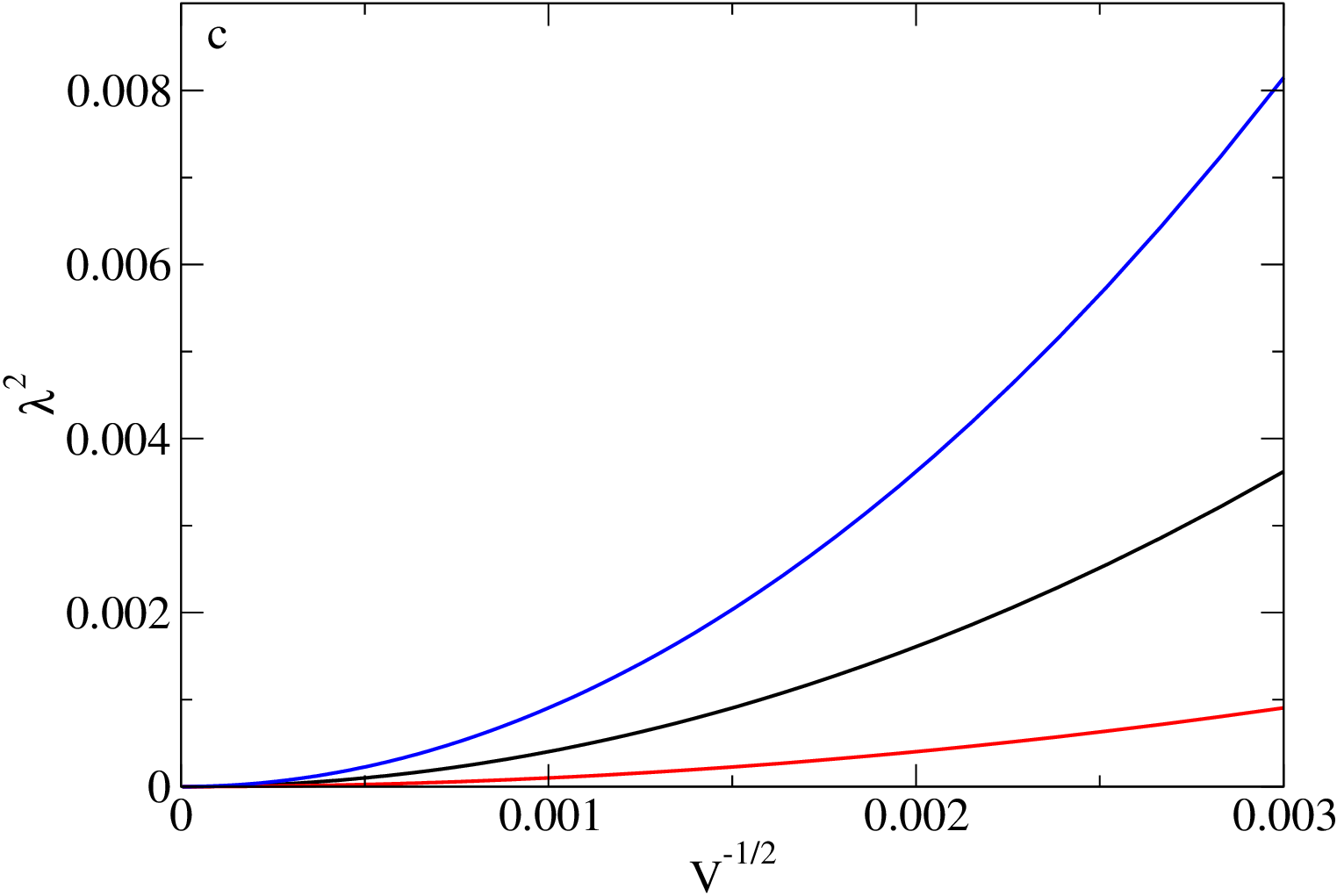}
\caption{
a: Dependence of $\omega_k$ on $V^{-1/2}$.
b: Dependence of $\sqrt{\omega_k}\lambda$ on $V^{-1/2}$.
c: Dependence of $\lambda^2$ on $V^{-1/2}$.
$L_x=L_y=L_z$ and the smallest volume is $L_x=40$.}
\label{cav1}
\end{figure}

The Fock space in Eq. \eqref{fock} can quickly  make the dimension of the
Hamiltonian (in Eq. \eqref{hamil}) is unmanageable and truncation is
needed. We will introduce a maximum frequency  $\omega_{cut}$ and
only $\omega_k<\omega_{cut}$ will be considered. Additionally, we
have also tested that for coupling strengths up to $\lambda$=0.1, 
the occupation probability of the Fock states with
$\sum_{k=1}^M n_k>1$ are negligibly small ($< 10^{-8}$). Thus we will
only use the photon-less $\vert\boldsymbol{0}\rangle$ space and Fock spaces
where only one $n_i=1$ and the rest is zero. We will refer to the
latter as $\omega_i$ space where all photon spaces are empty except
for the $\omega_i$ space which has occupation number $n_i$=1.
\begin{figure}
\includegraphics[width=0.45\textwidth]{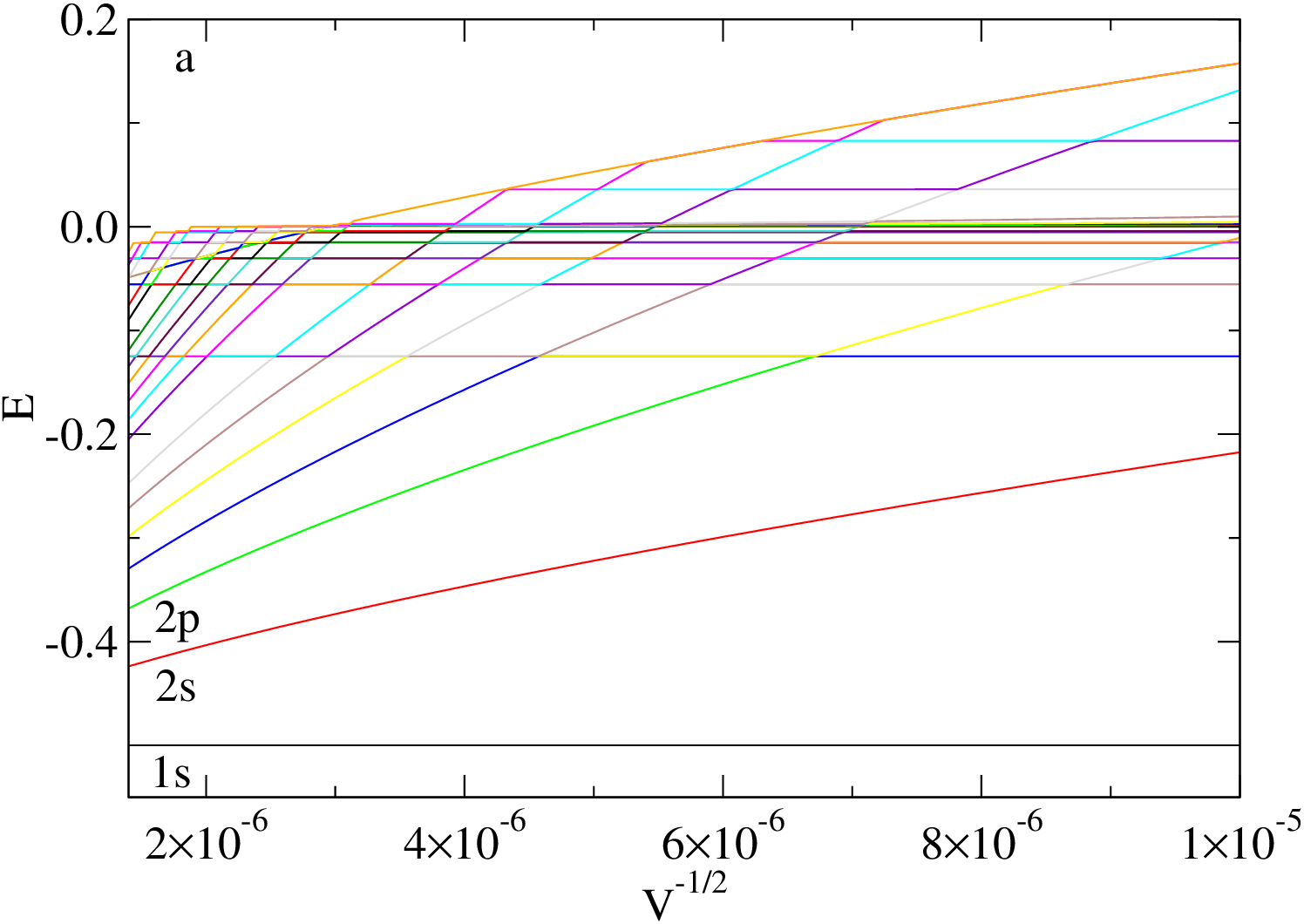}\\
\includegraphics[width=0.45\textwidth]{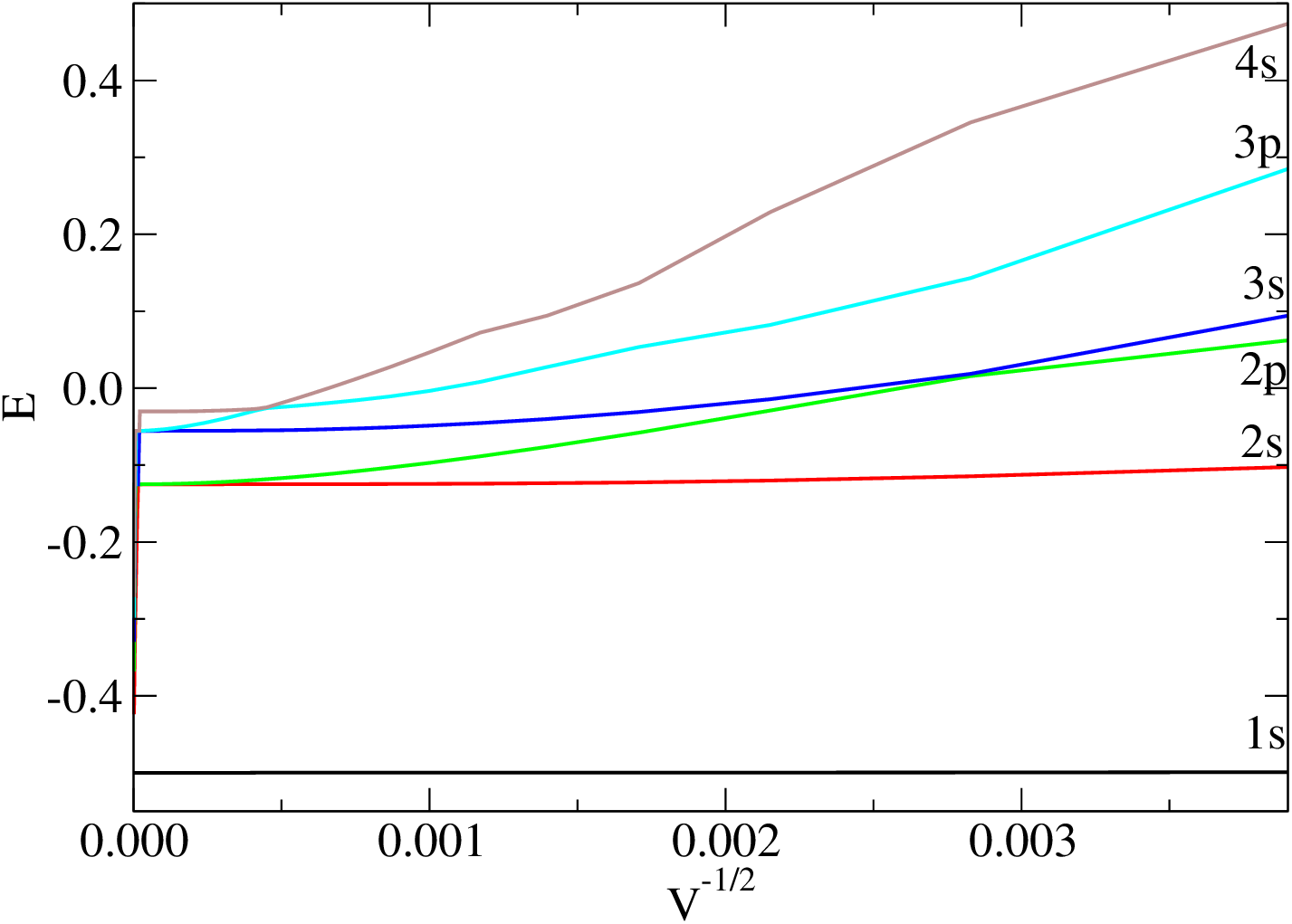}
\caption{
a: Dependence of the energy on V$^{-1/2}$ in the L=2000-8000 region.
b: Dependence of the energy on V$^{-1/2}$ in the L=40-8000
region.}
\label{cav2}
\end{figure}

To highlight the roles of different parts of the Hamiltonian we plot
them in Fig. \ref{cav1}. The figure shows the magnitude of the
different terms as a function of the cavity volume for an $L_x=L_y=L_z$ cavity. 
As Eq. \eqref{node} shows,  the frequency is proportional to $V^{-1/3}$
and $\lambda$ is proportional to $V^{-1/2}$. We will
use $V^{-1/2}$ to show the volume dependence in the following.
The $\omega_k$ frequencies change rapidly with decreasing 
the volume of the cavity (Fig. \ref{cav1}a). 
This term adds $n_k \omega_k$ energy to the total energy of the states containing photons 
but does not change the states besides the energy shift (see Eq.
\eqref{hamilm}). The
coupling term (Eq. \eqref{coupm}) is proportional to $\sqrt{\omega}\lambda$ 
and contributes significantly to the Hamiltonian (Fig. \ref{cav1}b). 
One can see that the sign of the coupling term 
can be positive or negative leading to cancellations.
The self-interaction term (the last term in Eq. \eqref{hamilm}) 
is proportional to $\lambda^2$ (Fig.\ref{cav1}c) and much smaller 
than the other terms.

\subsection{Energy Spectrum}
Fig. \ref{cav2} shows the change of the energy of the H atom  as the function of
the volume of the cavity. The Fig. \ref{cav2}a figure zooms in the region where the
cavity is large and Fig.\ref{cav2}b shows the full change when the
cavity size is decreased to $L_x=L_y=L_z=40$. In the 
larger cavity ($L_x=L_y=L_z=2000,{\ldots},8000$)
the energy levels are raising quickly  due to the strong effect of the
$\omega_kn_k$ term in Eq. \eqref{hamilm}. The hydrogenic states
(1s,2s,2p,{\ldots}) are coupled to each $\omega_k$ state. The lowest 
line in Fig. \ref{cav2} is the ground state coupled to the
$\vert\boldsymbol{0}\rangle$ photon-less state. The first excited state 
is the ground state of the Hydrogen atom coupled to the $\omega_1$
($|100{\ldots}{\ldots}\rangle$) state. The second excited state starts
as a hydrogen ground state coupled to the $\omega_2$ state but rises
to reach the energy of the 2s state of the Hydrogen coupled to the 
$\vert\boldsymbol{0}\rangle$ photon-less state. The coupling to any other
photon states is negligibly small. When the energy of the
two states get close to each other an avoided crossing occurs (this
will be discussed later) and the second excited state becomes a 2s state
coupled strongly to the photon-less state but with significant coupling 
to all other photon states. The entanglement with other photon spaces
increases by increasing the coupling strength (decreasing the
volume). Other higher excited states behave similarly, except that not
only the photon spaces but the hydrogen eigenstates can also be coupled
to each other with increasing the coupling strength. 

Fig.\ref{cav2}b shows the change of the lowest energy levels of the
Hydrogen when the cavity volume is decreased to $L_x=L_y=L_z=40$. 
The state labels (1s,2s,2p{\ldots}) are assigned based on the initial
energy and angular momentum (when the coupling was very weak at the large
volume limit). As the coupling increases these states become entangled
states of hydrogenic and photonic states; each state is a linear
combination of several coupled matter-photon states.
The change of energy of the 1s state
is small but it is coupled with all photon spaces. The 2s and 2p state
start as degenerate states but the energy of the $2p$ state changes
much more due to the dipole coupling. The figure also shows the
avoided crossing between the 2p-3s and the 3p-4s states.
The energy of higher excited states strongly depends on the light-matter coupling
(cavity volume). 

\begin{figure}
\includegraphics[width=0.45\textwidth]{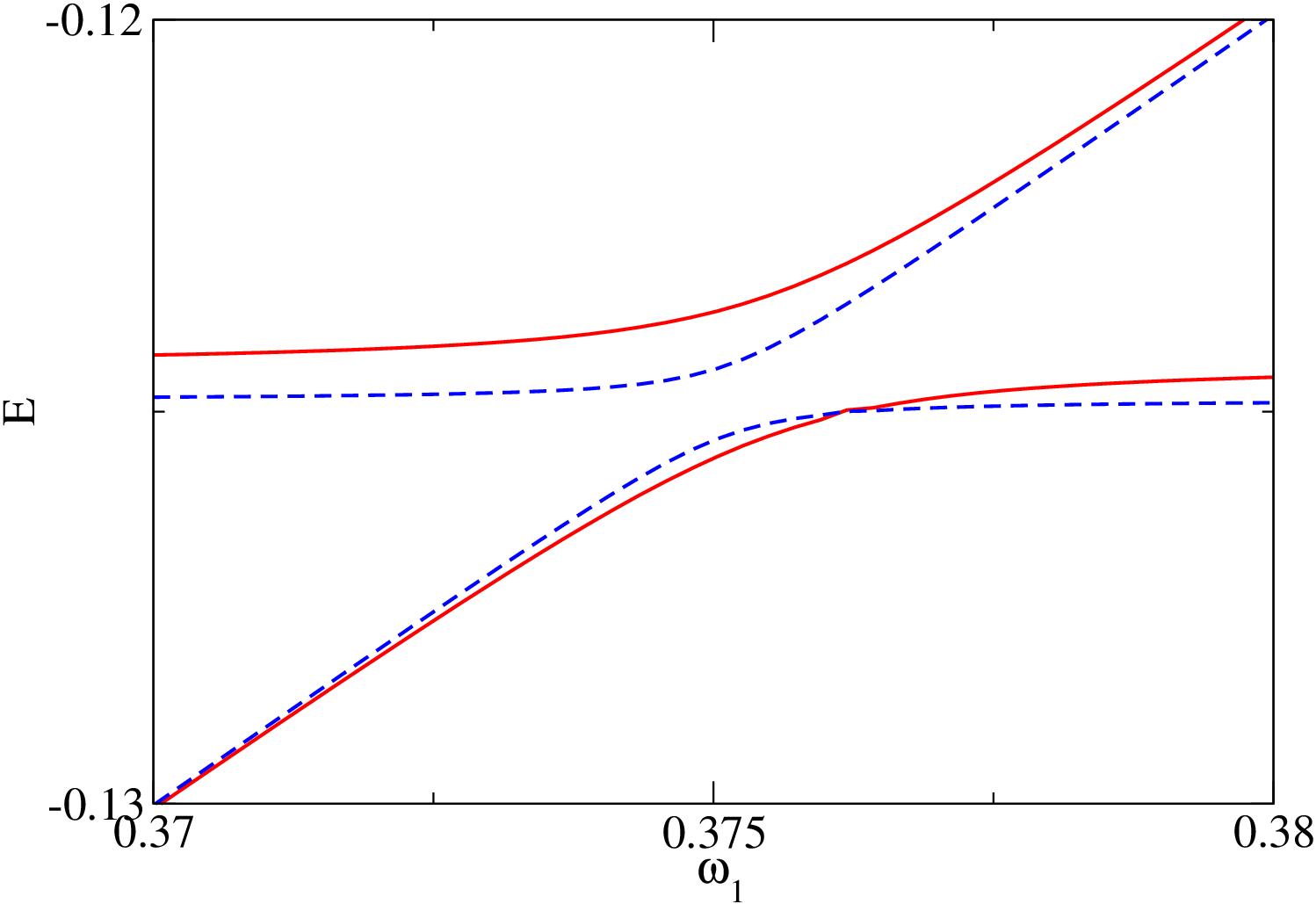}
\caption{Avoided level crossing as a function of $\omega_1$. The lowest
frequency was tuned  by changing $L_y=L_z$ around 1623 for $L_x=20$ (solid line)
and $L_x=80$ (dashed line). }
\label{ale}
\end{figure}

Fig.\ref{ale} zooms in an avoided crossing region. Avoided crossing
\cite{Heiss_1990,https://doi.org/10.1002/prop.201200062}
occurs when two energy levels approach each other but due to coupling
between them the energy of the two states cannot be degenerated and will
move on hyperbolic paths. The distance between the two paths is
proportional to the coupling \cite{PhysRevE.64.036213}. Fig.\ref{ale}
shows the change in the energy of the 1s and 2p states. Both of these
states are coupled to the $\vert\boldsymbol{0}\rangle$  space and the
$\omega_1$ space which has one photon with frequency $\omega_1$ and
all other frequencies are unoccupied. We emphasize that all matter
states and all photon states are coupled, and naming the matter states
1s or 2p means that the dominant component of the state is a 1s or a 2p
state. The dominant components are strongly coupled to the
$\vert\boldsymbol{0}\rangle$  space and the  $\omega_1$ photon states
and the probability of other photon states are less than 0.001. 

To show the avoided crossing we change the lowest cavity frequency
$\omega_1$ at around the 1s-2p excitation energy
by fixing $L_x$ and changing $L_y=L_z$ around 1623.
Fig.\ref{ale} shows that the energy of the lower 1s state nearly linearly
increases up to 0.375 due to the $\omega_k$ term in Eq. \eqref{hamil} 
(see also Fig. \ref{cav1}a) and when it reaches its closest distance to the 2p
states its slope decreases. The opposite is true for the 2p state, it
starts flat and increases linearly after 0.375. The closest distance is
not exactly at 0.375 because the energy difference between the two
states is slightly different in the cavity. 
Fig. \ref{ale} shows the avoided crossing for two different cavity
volumes. The gap increases by decreasing the cavity volume
(increasing the coupling). As there are no states in 
the gap there is no light absorption for
that energy (this is sometimes referred to as electromagnetically induced
transparency). This gap is also called as Rabi splitting. 

\begin{figure}
\includegraphics[width=0.45\textwidth]{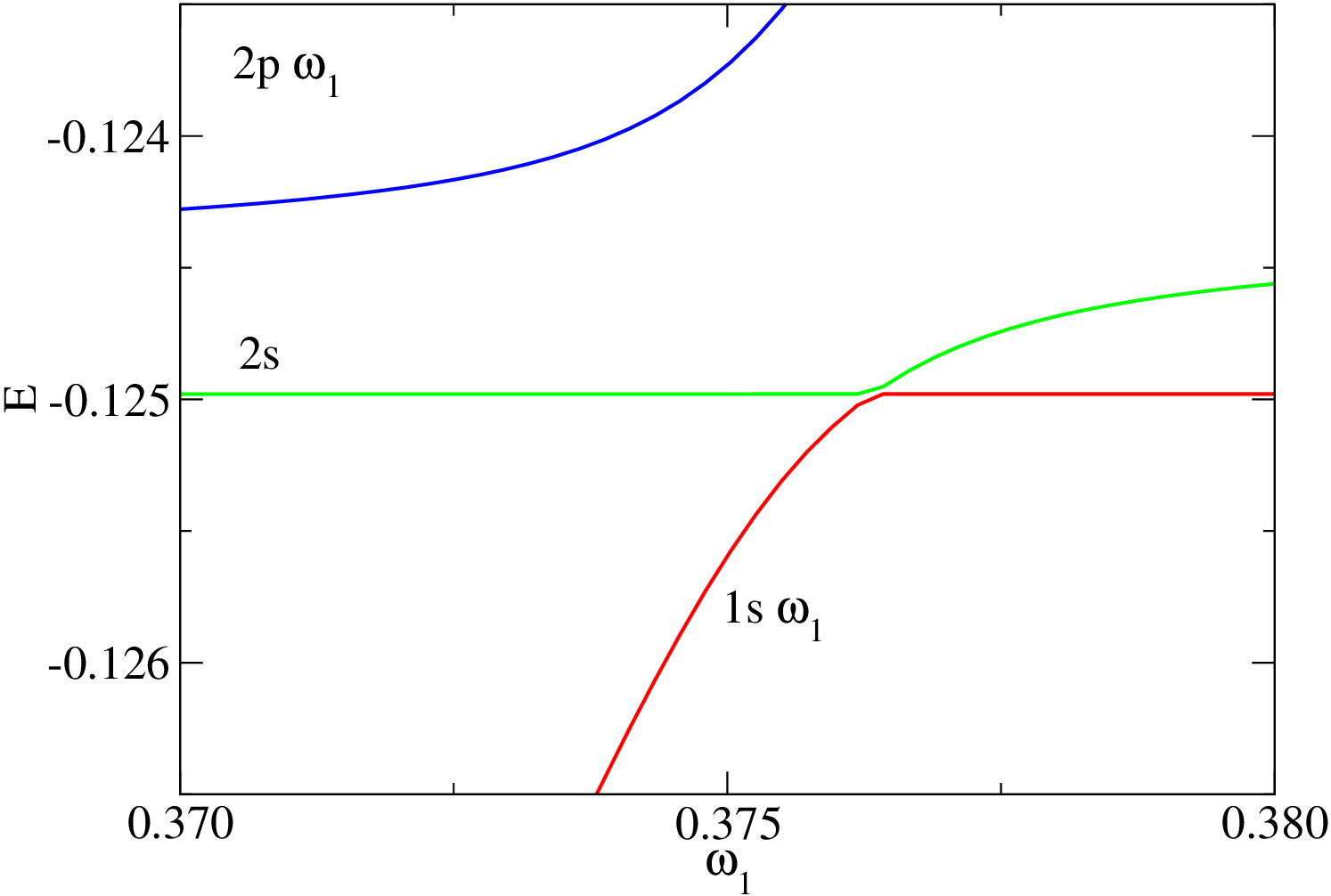}
\caption{Avoided level crossing as a function of $\omega_1$. The lines
are labeled by the dominant matter state ($L_x=20$).}
\label{ale1}
\end{figure}

In this system, the avoided level crossing occurs not only between two
states but between several states. Fig. \ref{ale1} (same as Fig.
\ref{ale} but now the $2s$ state is also included) shows 
the change of the energy levels with $\omega_1$. The figure shows
that in addition to the avoided crossing between the 1s and 2p states,
the 1s state also avoids the 2s state. The gap between the 1s and 2s
state is very small because the coupling is small. The 1s and 2s state
is not directly coupled (the dipole matrix element between these state
is zero), but they are both coupled to other states, e.g. the 2p state.

\begin{figure}
\includegraphics[width=0.45\textwidth]{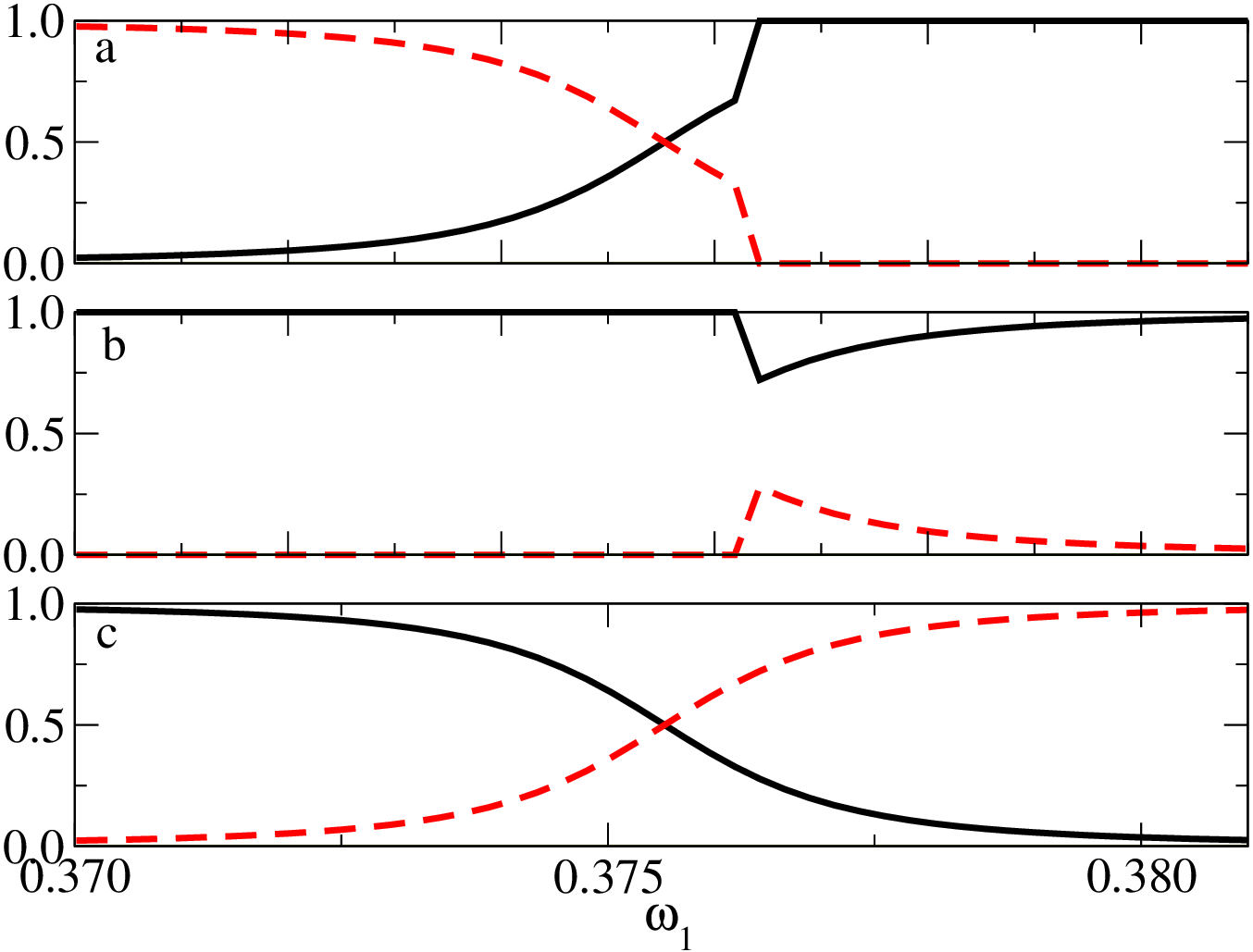}
\caption{Photon occupation probabilities in 
the $\vert\boldsymbol{0}\rangle$ space (solid line) and in the
$\omega_1$ space which has one photon with frequency $\omega_1$ and
all other frequencies are unoccupied (dashed line) for the 1s (a), 2s
(b) and 2p (c) states.}
\label{ale2}
\end{figure}

Fig. \ref{ale2} shows how the photon occupation probabilities change
at the avoided crossing. The 1s state (Fig.\ref{ale2}a) is first coupled
to the $\omega_1$ state (that is why its energy increases linearly
with $\omega_1$ on Fig.\ref{ale1}) and around the avoided crossing it is coupled to the
$\omega_1$ and $\vert\boldsymbol{0}\rangle$ states and after the avoided crossing it is
fully coupled to $\vert\boldsymbol{0}\rangle$. The 2s state
(Fig.\ref{ale2}b) is coupled to the $\vert\boldsymbol{0}\rangle$
state and around the avoided crossing it coupled both to
$\vert\boldsymbol{0}\rangle$ and $\omega_1$ and returns to
$\vert\boldsymbol{0}\rangle$ after that. The 2p state (Fig.
\ref{ale2}c) is first coupled to $\vert\boldsymbol{0}\rangle$ only, it
is coupled to both $\vert\boldsymbol{0}\rangle$ and $\omega_1$ around
the avoided crossing  and
finally it is only coupled to $\omega_1$. The 1s and 2p states switch
coupling from $\vert\boldsymbol{0}\rangle$ to $\omega_1$ between each
other at the avoided crossing. The 2p state coupled to $\omega_1$
continue to rise in energy and will reach a state above it leading to
another avoided crossing and switching photon states.

\subsection{Optical absorption spectrum}

In the cases we have discussed so far the energy spectrum was
calculated by diagonalizing large sparse matrices with dimensions up
to 10000. This is possible
for a one-electron system, but not  for larger atoms or molecules.
In Ref. \cite{10.1063/5.0123909} we have introduced a QED time-dependent
density functional theory (QED-TDDFT) calculations which solve a PF Hamiltonian using a
product basis of a real space grid and a single Fock space. The 
present approach can be used to extend the QED-TDDFT to multiply
photon spaces, but diagonalization is not possible in the
QED-TDDFT case. To avoid diagonalization one can determine the excited
states by calculating the optical absorption spectrum by solution of the
time-dependent Schr\"odinger equation by time propagation. In the
calculation, we start with a delta-kick perturbation (see eq. \ref{kick})
which excites all states in the spectrum. 
In the following, we show examples of how the absorption spectrum
changes in a cavity. 

\begin{figure}
\includegraphics[width=0.45\textwidth]{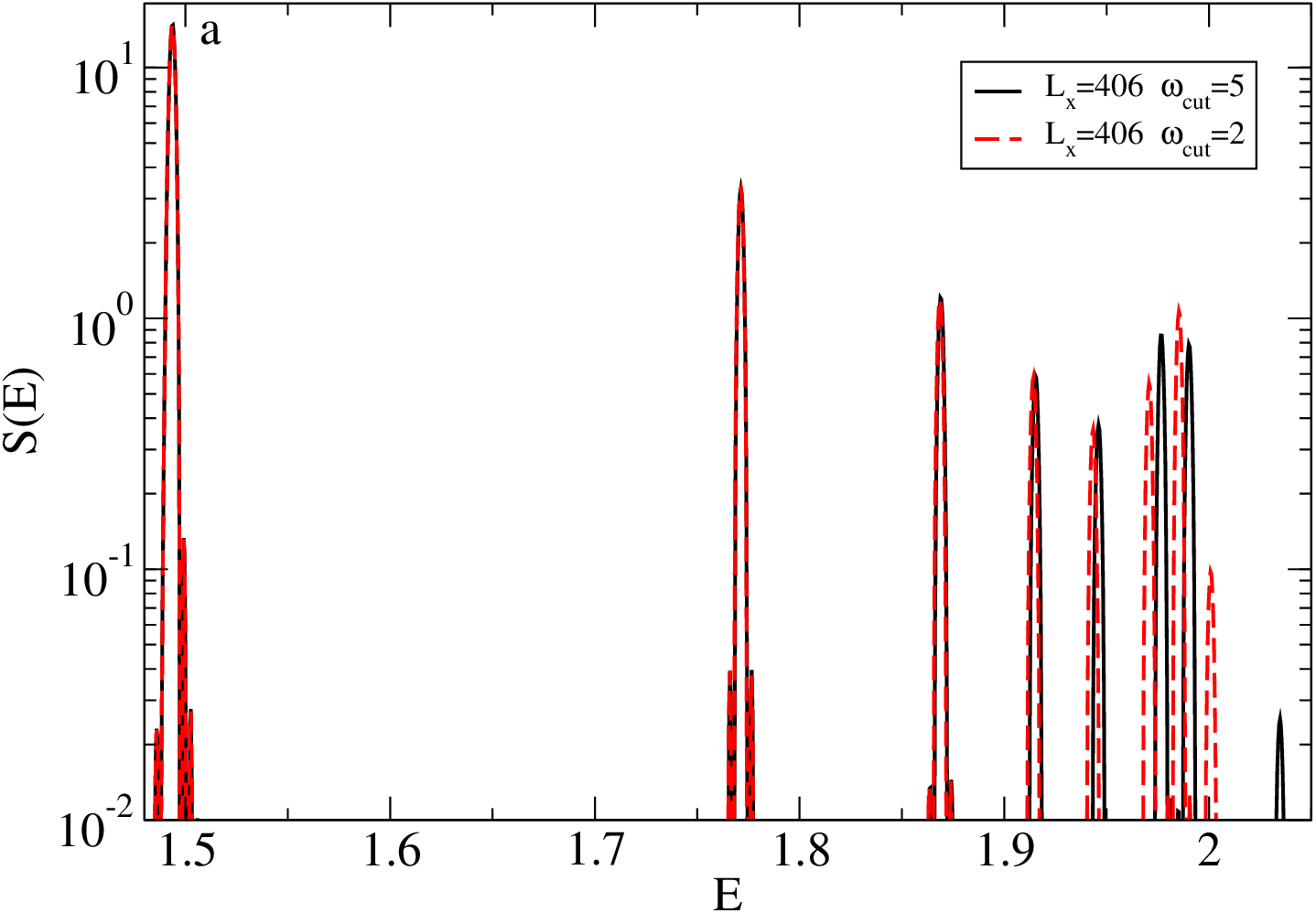}\\
\includegraphics[width=0.45\textwidth]{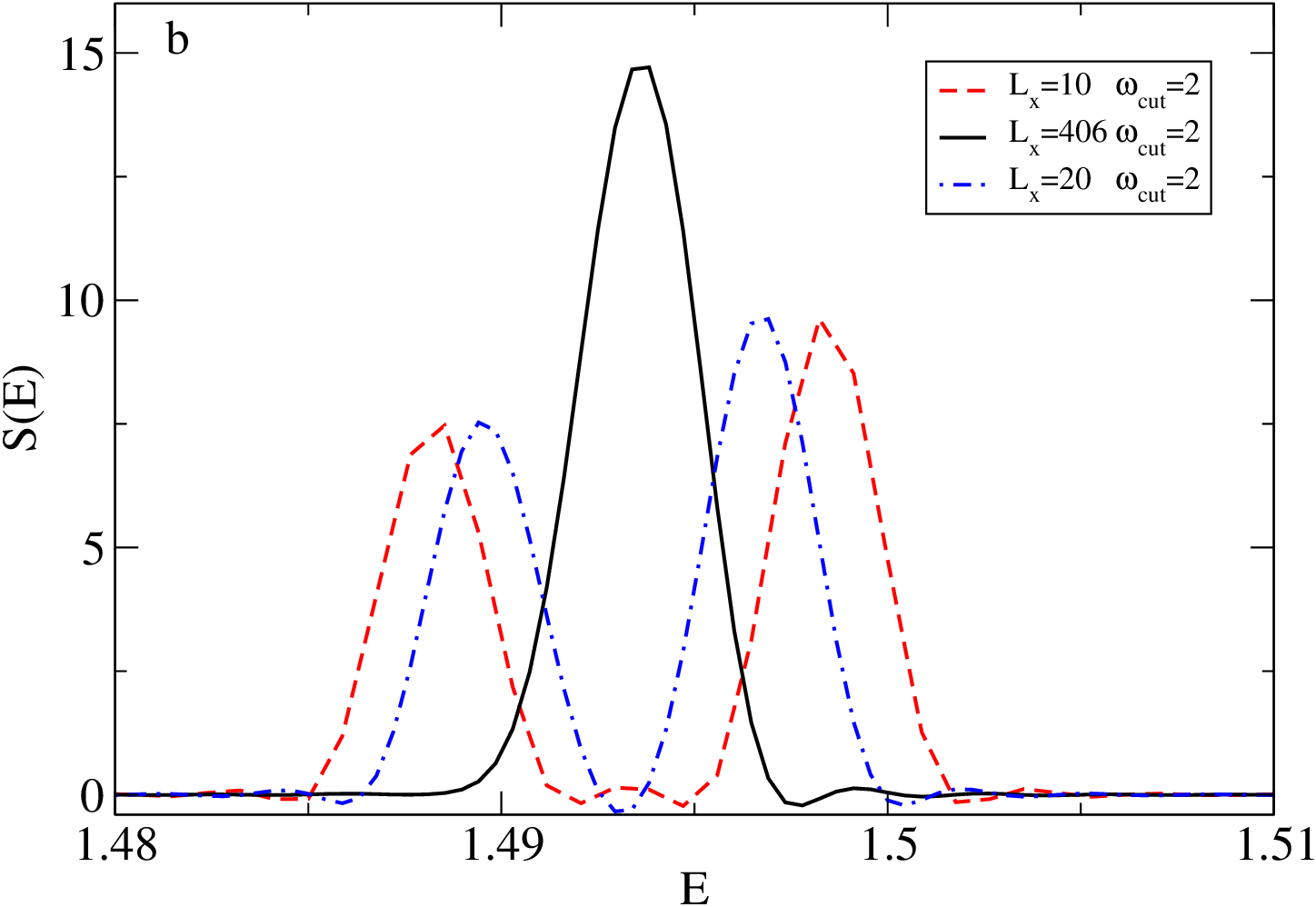}\\
\includegraphics[width=0.45\textwidth]{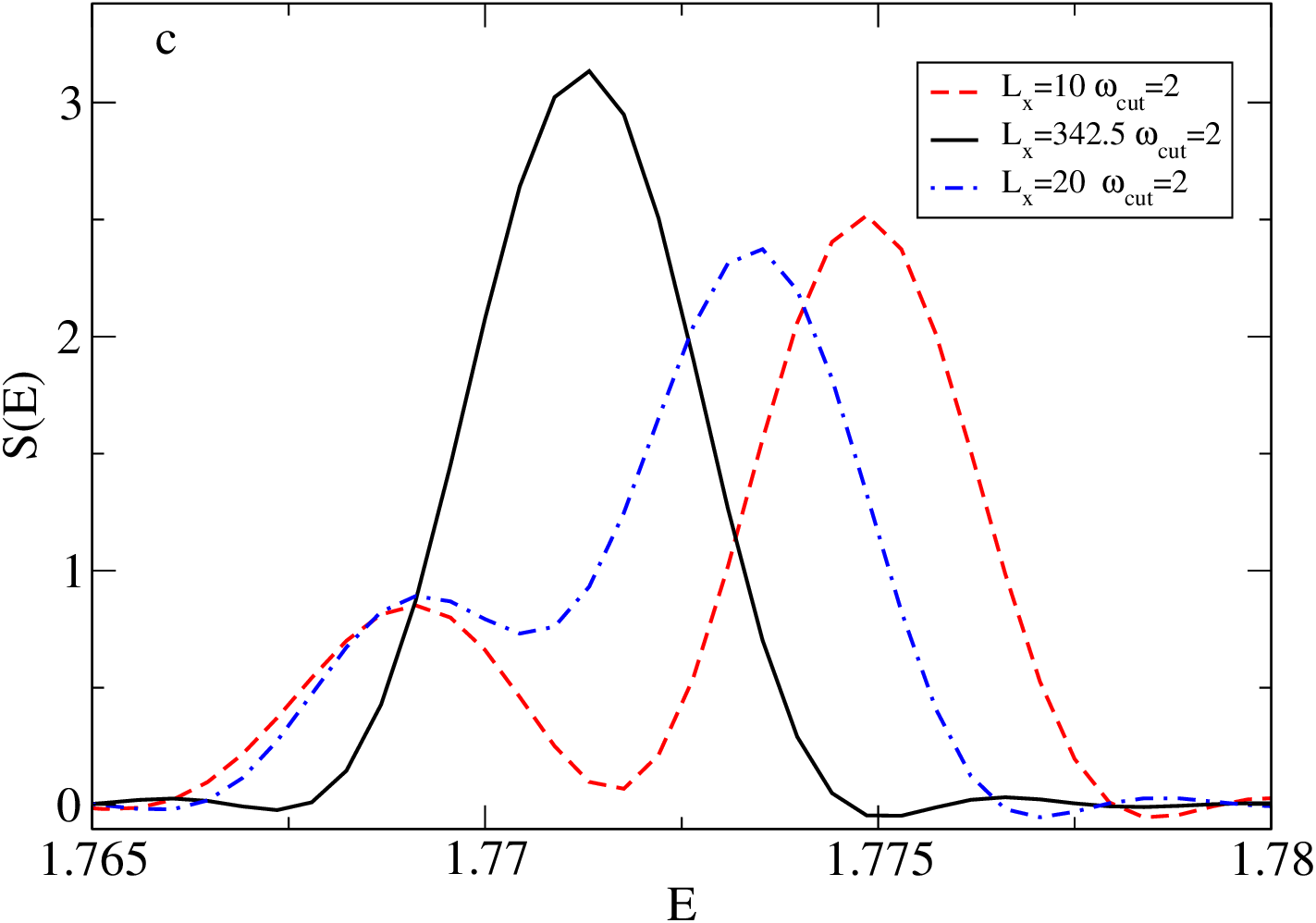}
\caption{a: Absorption spectrum for different values of $\omega_{cut}$.
The cavity size is $L_x=L_y=L_z=406$ and
$
\omega_1=1.5, \omega_2=2.6, \omega_3=3.35,
\omega_4=3.97, \omega_5=4.5, \omega_6=4.97,
\lambda_1=l,\lambda_2=-l/2,\lambda_3=-l,
\lambda_4=l,\lambda_5=l,\lambda_6=-l/2$ and
$l=0.0025$.
b: Absorption spectrum showing the splitting of the $E=1.5$ peak.
The cavity size is $L_y=L_z=406$ and $L_x$ is varied. For $L_x=20$
$\omega_1=1.5$ and $\lambda=0.01$, for $L_x=10$ 
$\omega_1=1.5$ and $\lambda=0.015$.
c: Absorption spectrum showing the splitting of the $E=2-2/9=1.778$ peak.
The cavity size is $L_y=L_z=342.5$ and $L_x$ is varied.
$\omega_1=1.777$ for all $L_x$ values. For 
$L_x=342.5$, $\lambda_1=0.003$,
$L_x=20$, $\lambda_1=0.0131$, 
$L_x=10$, $\lambda_1=0.0185$}
\label{split}
\end{figure}

In Fig.\ref{split}a we show the absorption spectrum peaks for a He$^+$
ion in a rectangular cavity for two different frequency cutoffs. The figure shows
that the low energy peaks are not sensitive to the cutoff value and
only the peaks close to the $E$=2 (continuum) threshold change
somewhat. The cavity size is chosen so that the lowest cavity frequency 
is equal to the 1s to 2p transition energy ($\omega_1=1.5$). Using
a lower frequency cutoff significantly reduces the basis dimensions
and thus the  calculation time. 

Fig.\ref{split}b shows the Rabi splitting of the 2p excited state of
the He$^+$ ion in a cavity. The figure shows that when the volume of
the cavity is large ($L_x=L_y=L_z=406$) then there is only one peak
at the 1s 2p transition in the absorption spectrum. By decreasing 
$L_x$ to 20, $\lambda_1$ increases while the lowest cavity frequency 
remains the same ($\omega_1=1.5$, because the frequency is determined 
by $L_x=L_y=406$ choice). Due to the larger $\lambda_1$  
the absorption peak splits into two peaks. By increasing $\lambda_1$
further ($L_x=10$) gaps between the two peaks get larger
This is the same effect as the avoided crossing that we have shown in Fig.
\ref{ale}, but this time we show it using the absorption spectrum. 
The asymmetry of the peaks is due to the fact 
that the energy of the 1s 2p transition in the cavity is 
different from the (free space) analytical value and the difference
increases with the coupling term. 

Fig. \ref{split}c shows the same splitting for the 1s 3p transition of
the He$^+$ ion. The cavity size is chosen so that the lowest cavity
frequency corresponds to the 1s 3p transition energy ($E=2-2/9$).
The asymmetry of peaks is even larger in this case because the
transition  energy of the less tightly bound 3p state is more sensitive
to the coupling to the cavity. The gap between the peaks is smaller than
in the previous case and one would need stronger coupling to further 
increase the distance between the peaks.

\begin{figure}
\includegraphics[width=0.45\textwidth]{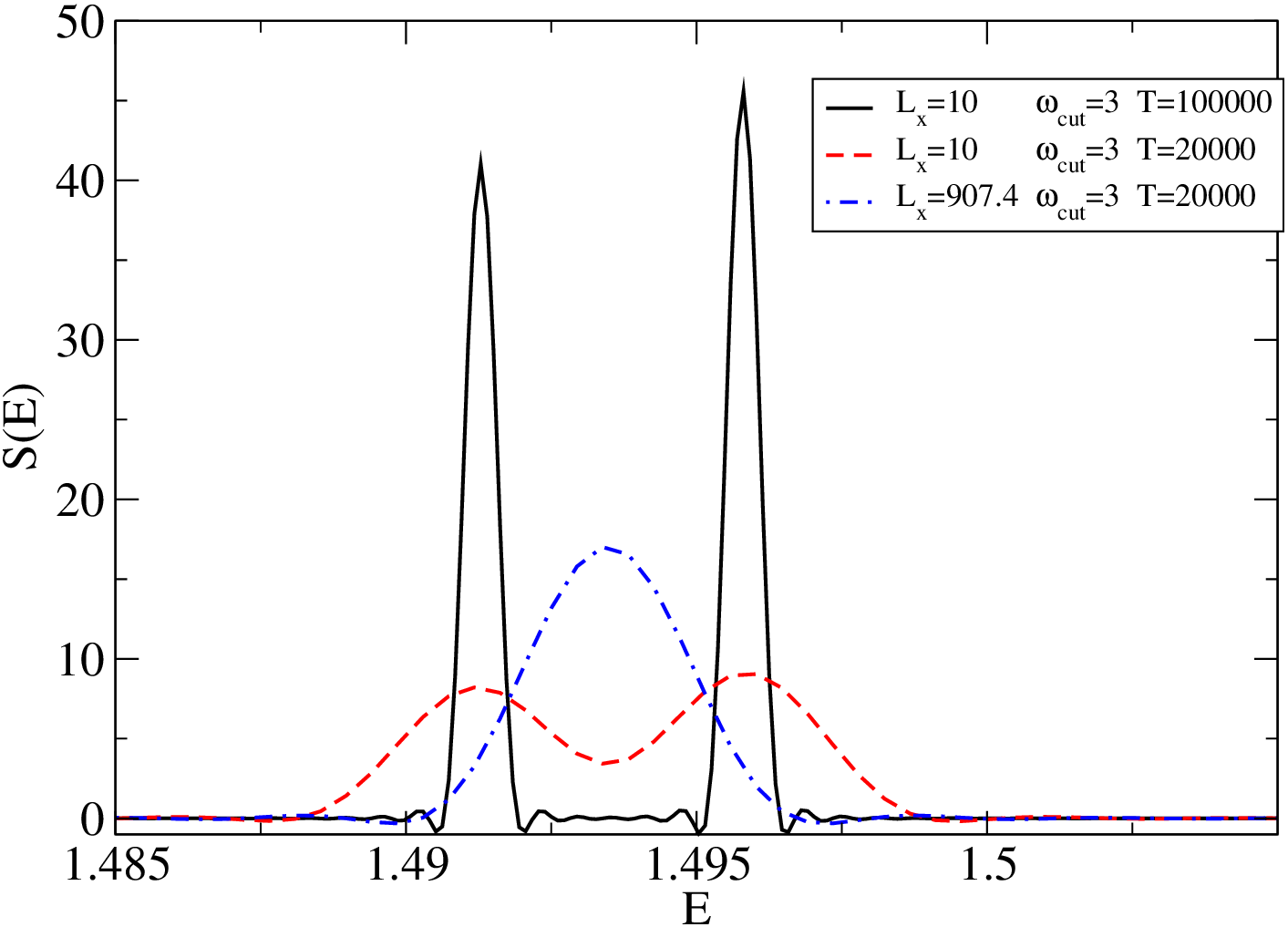}
\caption{Absorption spectrum showing the splitting of the $E=1.5$ peak
using the second photon mode. Cavity size $L_y=L_z=907.4$ and $L_x$
varied. 
$\omega_1=0.67, \omega_2=1.5, \omega_3=2.01, \omega_4=2.42,
\omega_5=2.77, \lambda_1=l,\lambda_2=-l,\lambda_3=l/2,
\lambda_4=l,\lambda_5=-l, l=0.007$}
\label{so}
\end{figure}

Fig. \ref{so} illustrates that any cavity mode can be used
to split a peak in the absorption spectrum. In this case the second cavity mode, $\omega_2=1.5$
(generated by adjusting the size of the cavity) is used to split  
the 1s 2p transition peak. The figure also shows that longer time
propagation leads the narrower peaks and after a very long time propagation 
Dirac-delta-like peaks would appear.

\subsection{High Harmonic Generation}
In atoms and small molecules, HHG can be understood as a three-step 
process \cite{PhysRevA.45.4998,PhysRevA.49.2117}. The laser removes
the electrons from the system with tunnel ionization and when the
direction of the driving field changes the electrons accelerate
backward recombining and scattering with the parent ion. This process 
leads to a pulse train of attosecond radiation bursts. By Fourier
transforming the temporal pulse train to the frequency space one gets a comb
of harmonic peaks. These peaks appear at frequencies that are the
integer multiples of the frequency of the exciting laser. These are not
the only intensity peaks that we expect to observe in the HHG spectrum.
Similarly to the absorption  spectrum case, the strong laser also
excites all states of the system, and peaks corresponding to the
transitions between different energy levels will also appear in the
HHG spectrum. 

\begin{figure}
\includegraphics[width=0.45\textwidth]{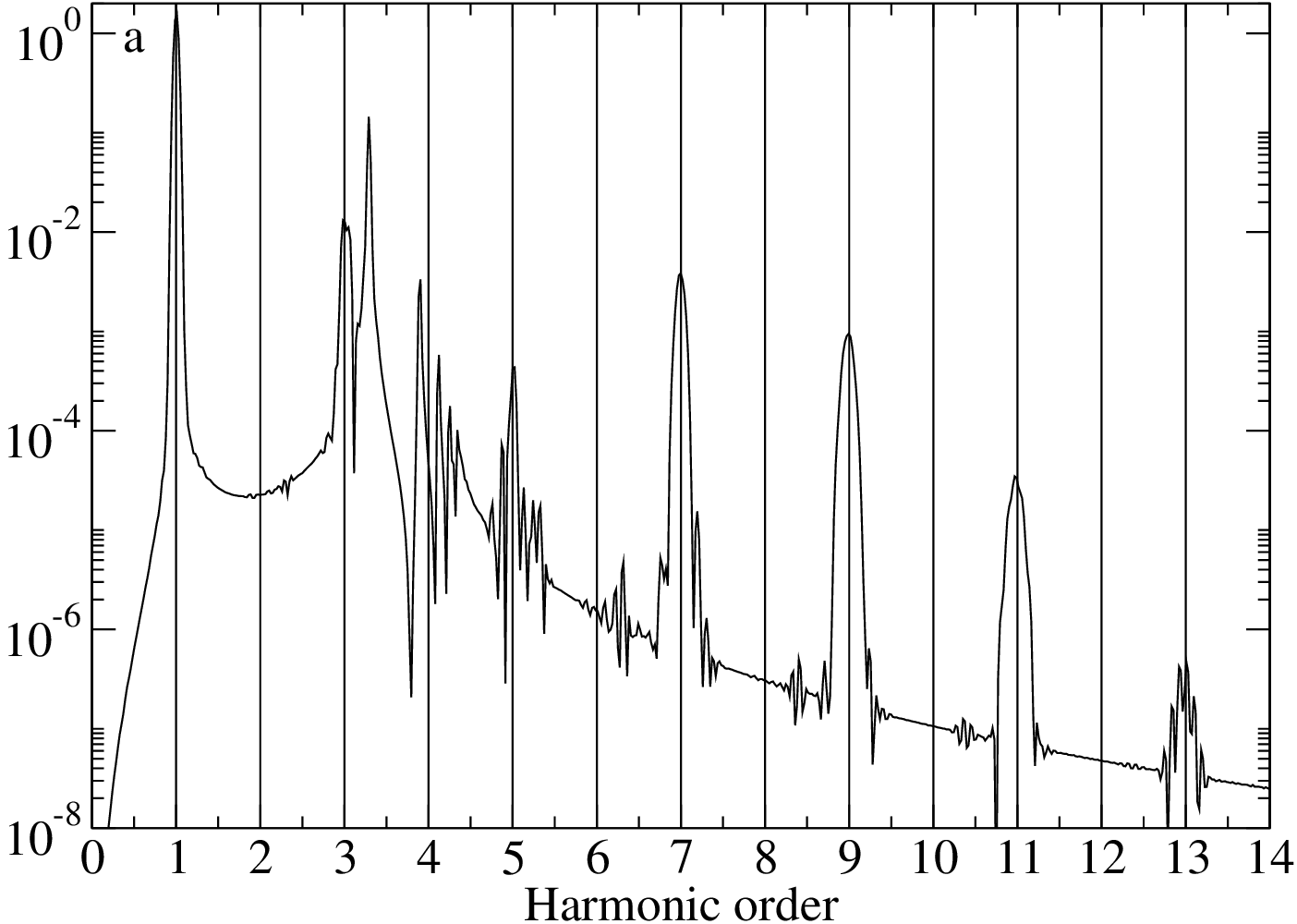}\\
\includegraphics[width=0.45\textwidth]{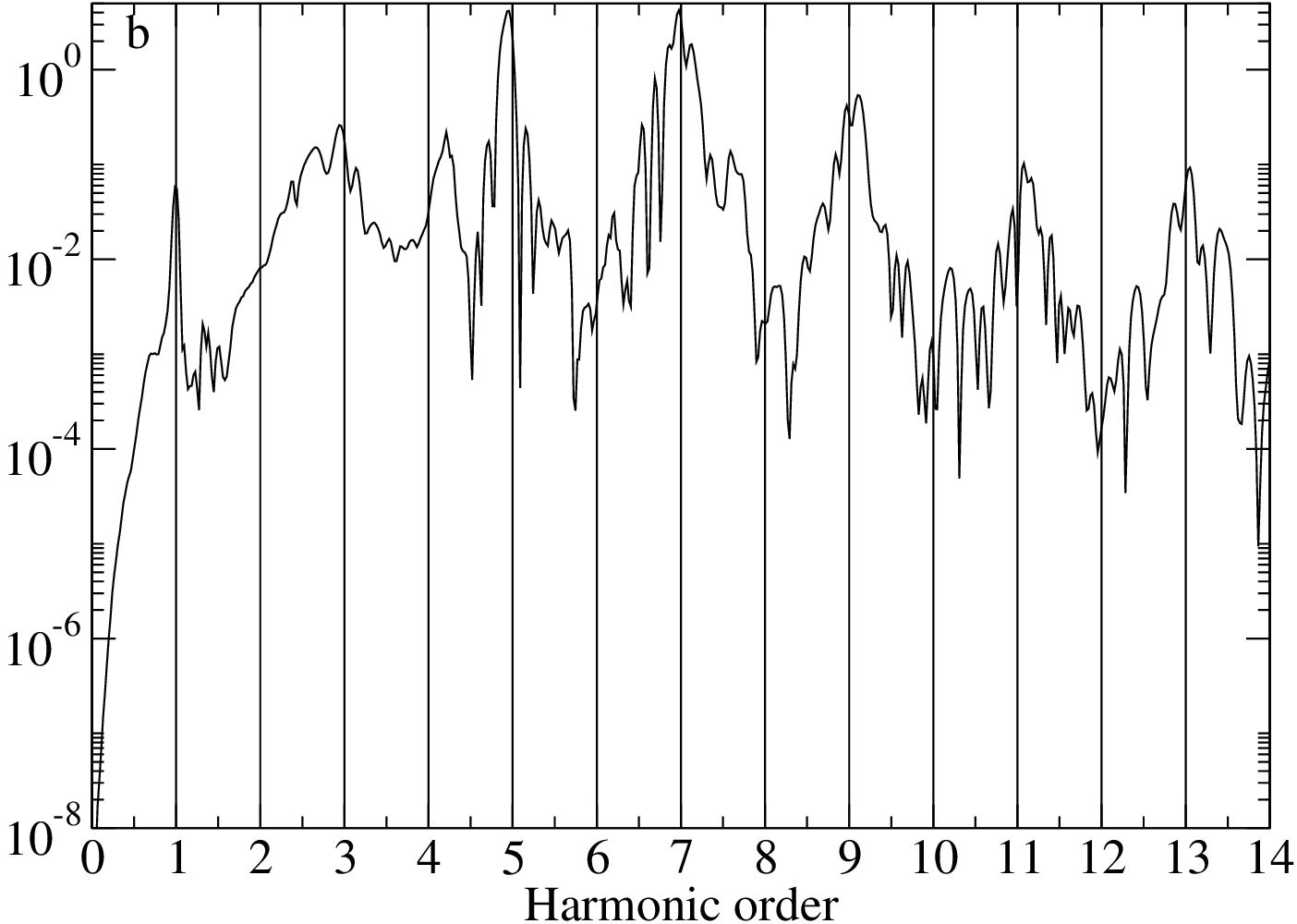}\\
\includegraphics[width=0.45\textwidth]{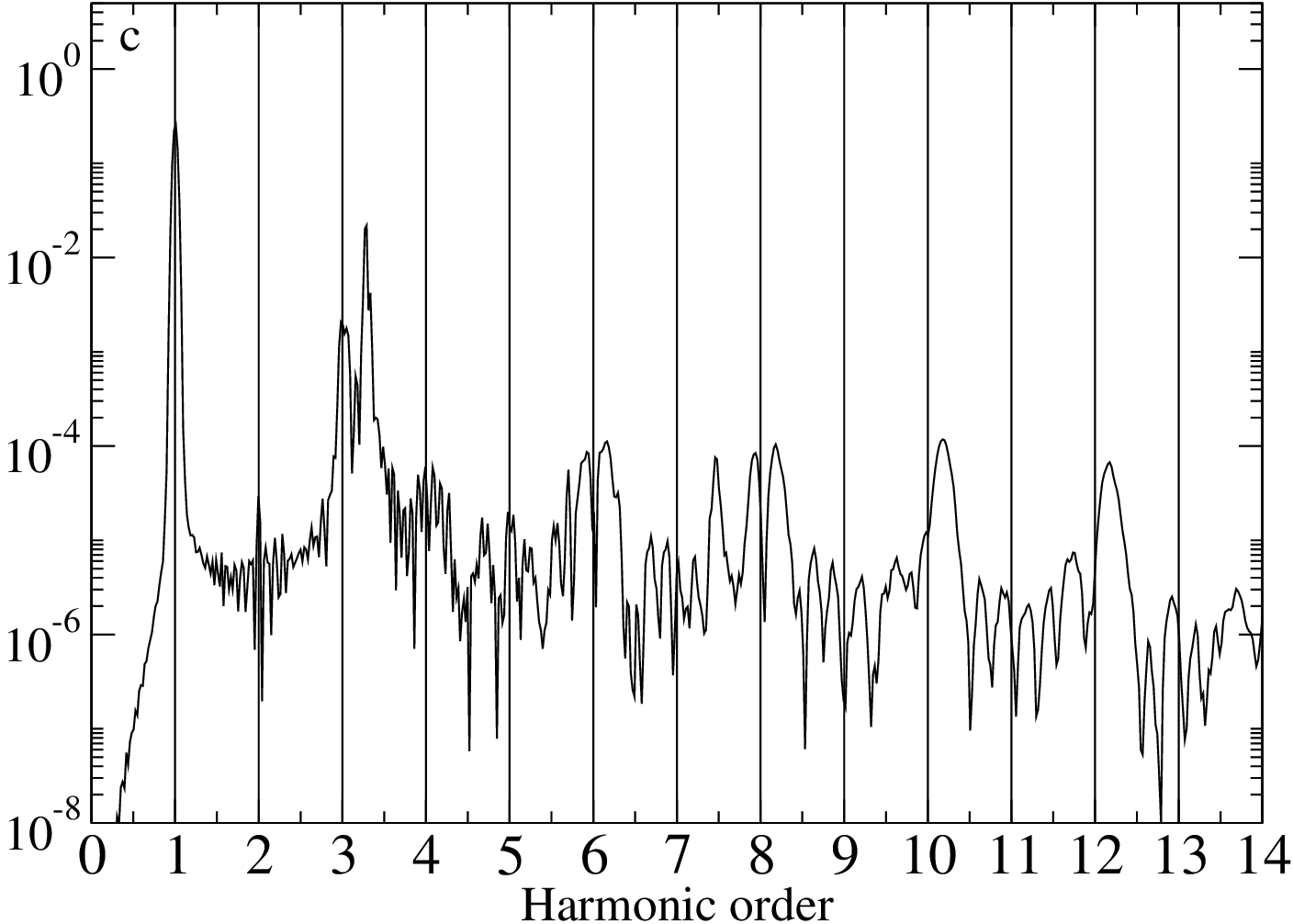}
\caption{
(a) HHG spectrum for free (no cavity) H atom.
(b) HHG spectrum for H atom in cavity $L_x=L_y=L_z=4000$.
(c) HHG spectrum for H atom in cavity $L_x=20,L_y=L_z=4000$. 
The
vertical axis is $I_h$ and the horizontal axis is $\omega\omega_h$.}
\label{hhg1}
\end{figure}

Fig.\ref{hhg1} shows the HHG spectrum of the H atom in free space and
in a cavity. The laser parameters are (see Eq. \eqref{laser}) 
$E_0=0.01$, $\omega_h=0.11395$, $t_0=15$ and $\tau=8.8$. The figure
shows the harmonic order which is defined as $\omega/\omega_h$ (see Eq.
\eqref{hhg}). In free space, due to the spherical symmetry, only the
odd harmonics are allowed. Fig.\ref{hhg1}a shows the 1st, 3rd, {\ldots}  
13th harmonics. The additional peaks are the excited states of the H
atom, for example, the 1s 2p transition energy 0.375 appears at
0.375/$\omega_h$=3.3, the 1s 3p transition is at 0.44444/$\omega_h$=3.9.  
Fig.\ref{hhg1}b shows the HHG spectrum in a large cavity. The intensity 
of the peaks is much larger in this case and the peaks also became 
wider due to the coupling with the cavity. Fig.\ref{hhg1}c shows the HHG
spectrum when $L_x$ is decreased to 20. The even harmonics are not
forbidden anymore and they appear in the spectrum, while the odd
harmonics are somewhat suppressed.

\begin{figure}
\includegraphics[width=0.45\textwidth]{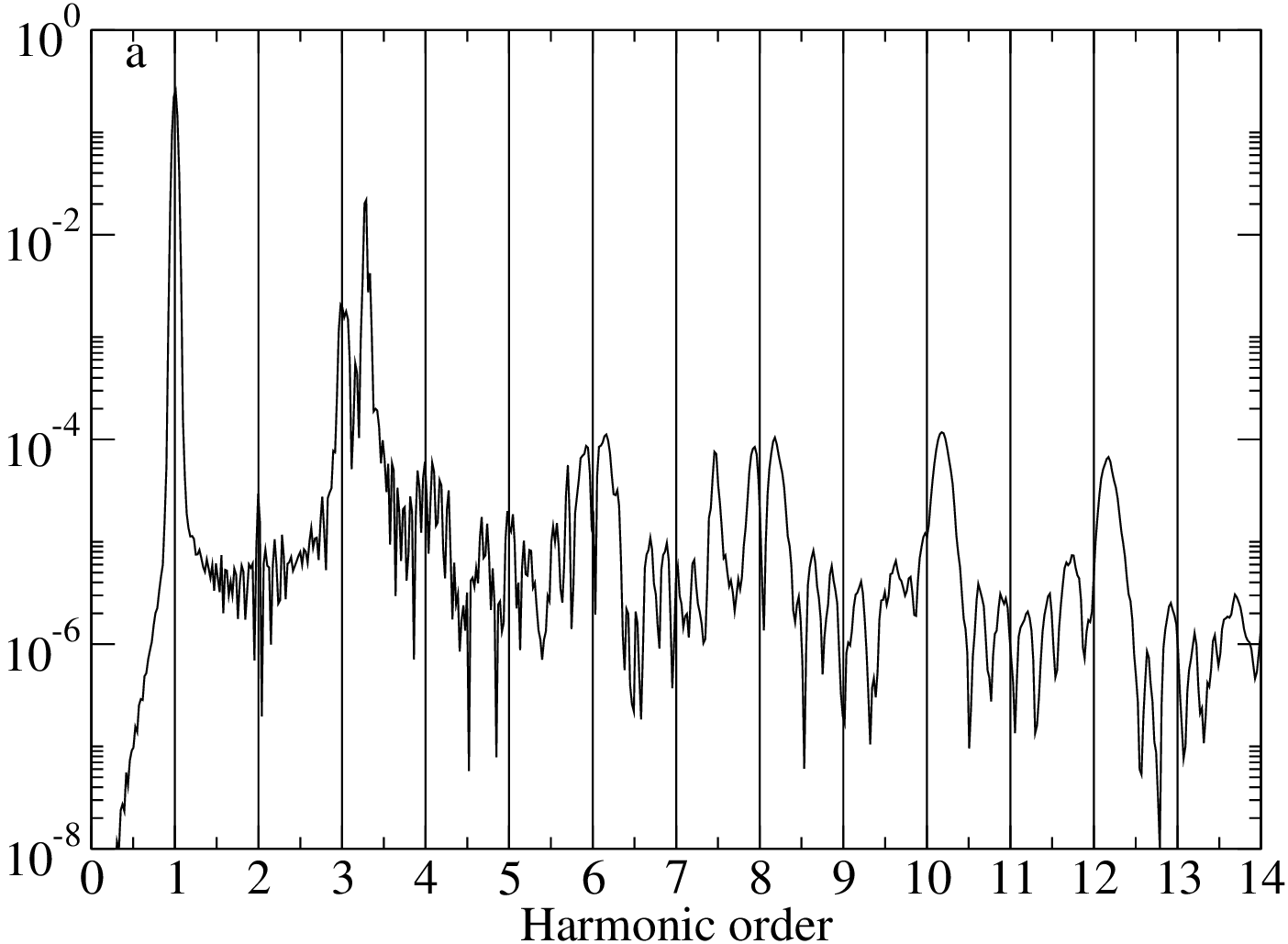}\\
\includegraphics[width=0.45\textwidth]{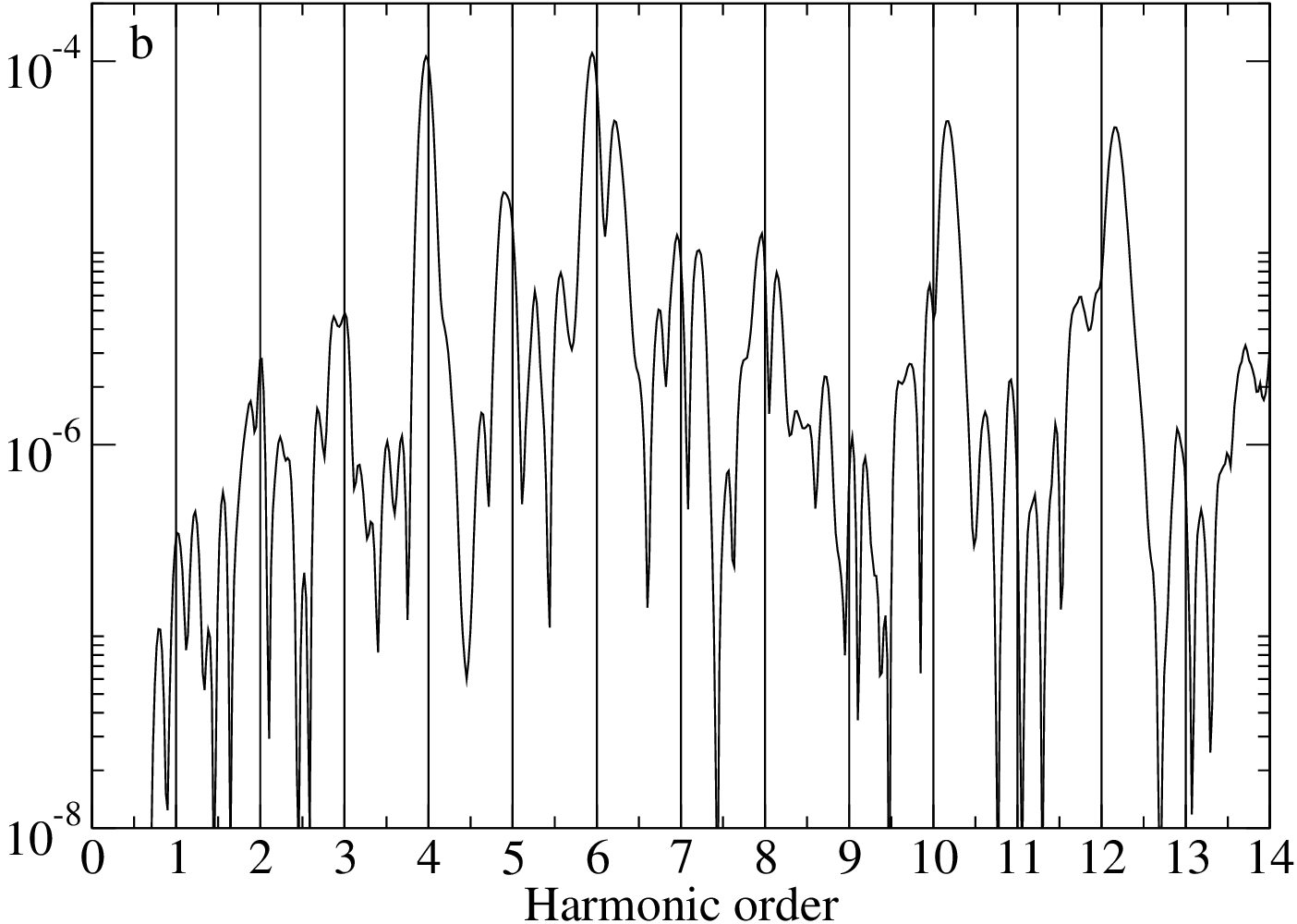}\\
\includegraphics[width=0.45\textwidth]{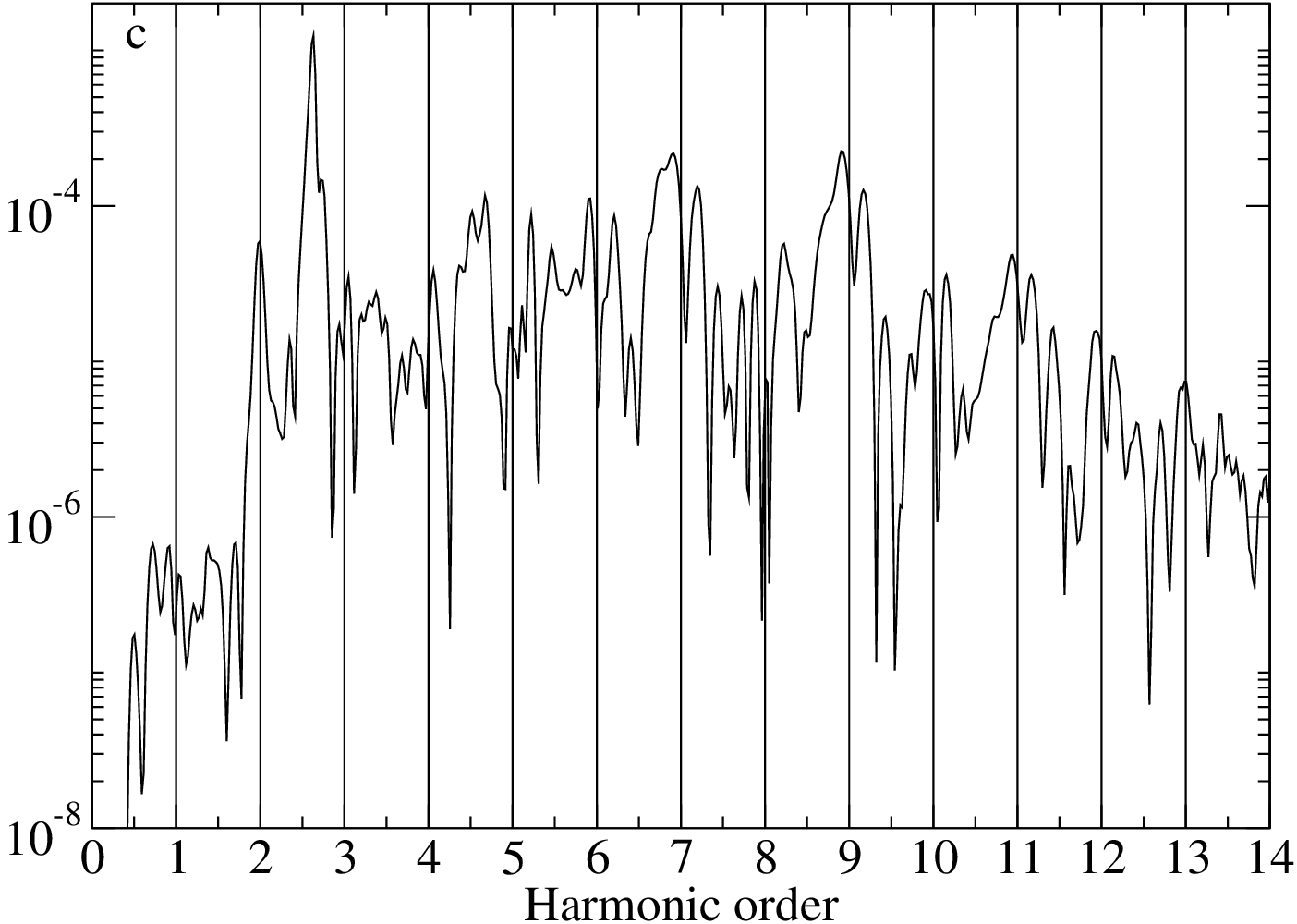}
\caption{
HHG spectrum for the first (a) second (b) and third states
(c) in a cavity of $L_x=20,L_y=L_z=2000$.
The vertical axis is $I_h$ and the horizontal axis is $\omega/\omega_h$.}
\label{hhg2}
\end{figure}

Fig.\ref{hhg1} shows the total HHG calculated using all photon spaces,
but using Eq. \eqref{pcon} one can calculate the HHG spectrum for the individual photon
spaces as well.
Fig.\ref{hhg2} shows the HHG spectrum for the
$\vert\boldsymbol{0}\rangle$, the $\omega_1$ and the $\omega_2$ photon
spaces. The figure illustrates, that HHG spectrum is present in every photon 
spaces, although the intensity in the $\vert\boldsymbol{0}\rangle$ is
larger. 

\begin{figure}
\includegraphics[width=0.45\textwidth]{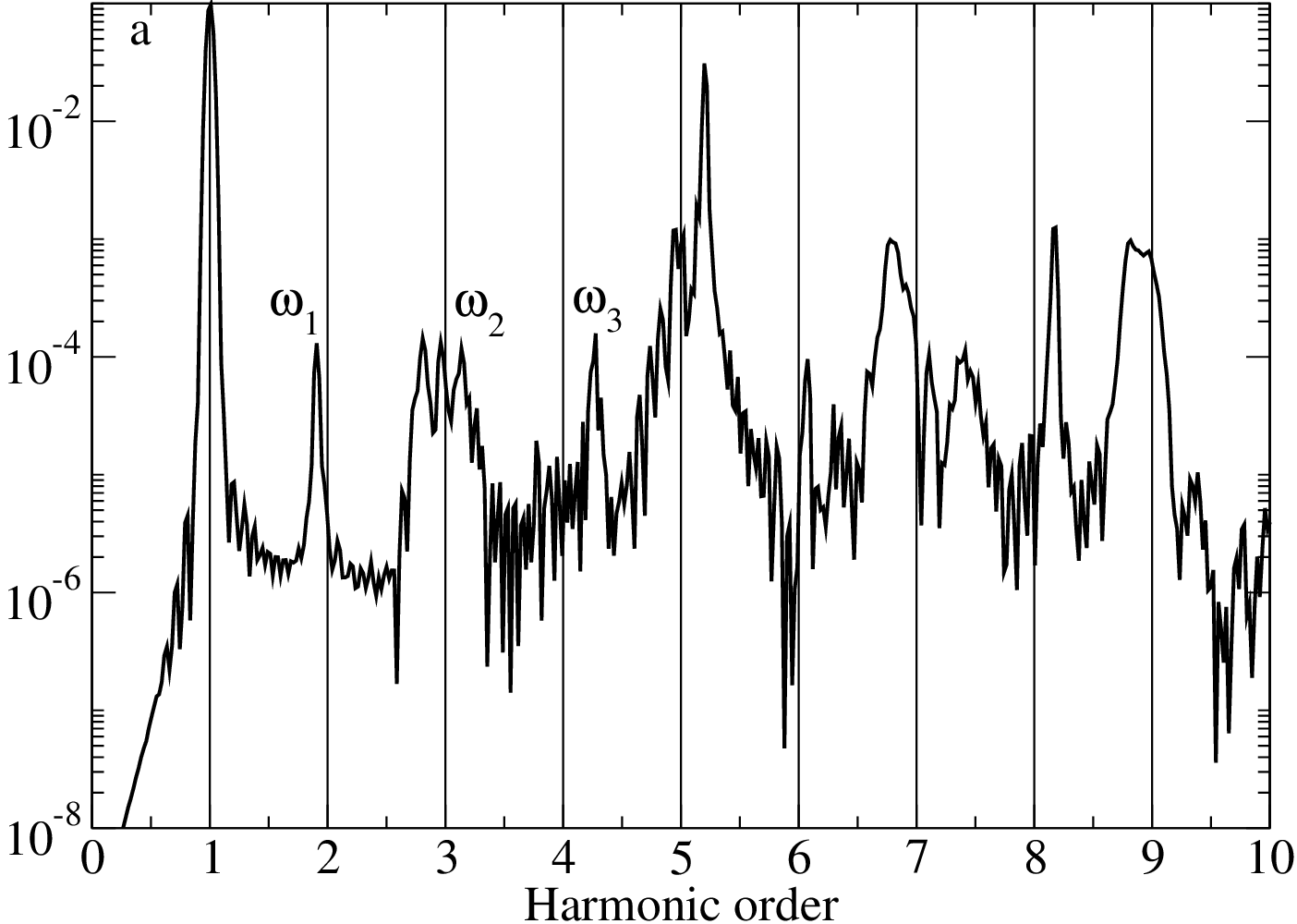}
\includegraphics[width=0.45\textwidth]{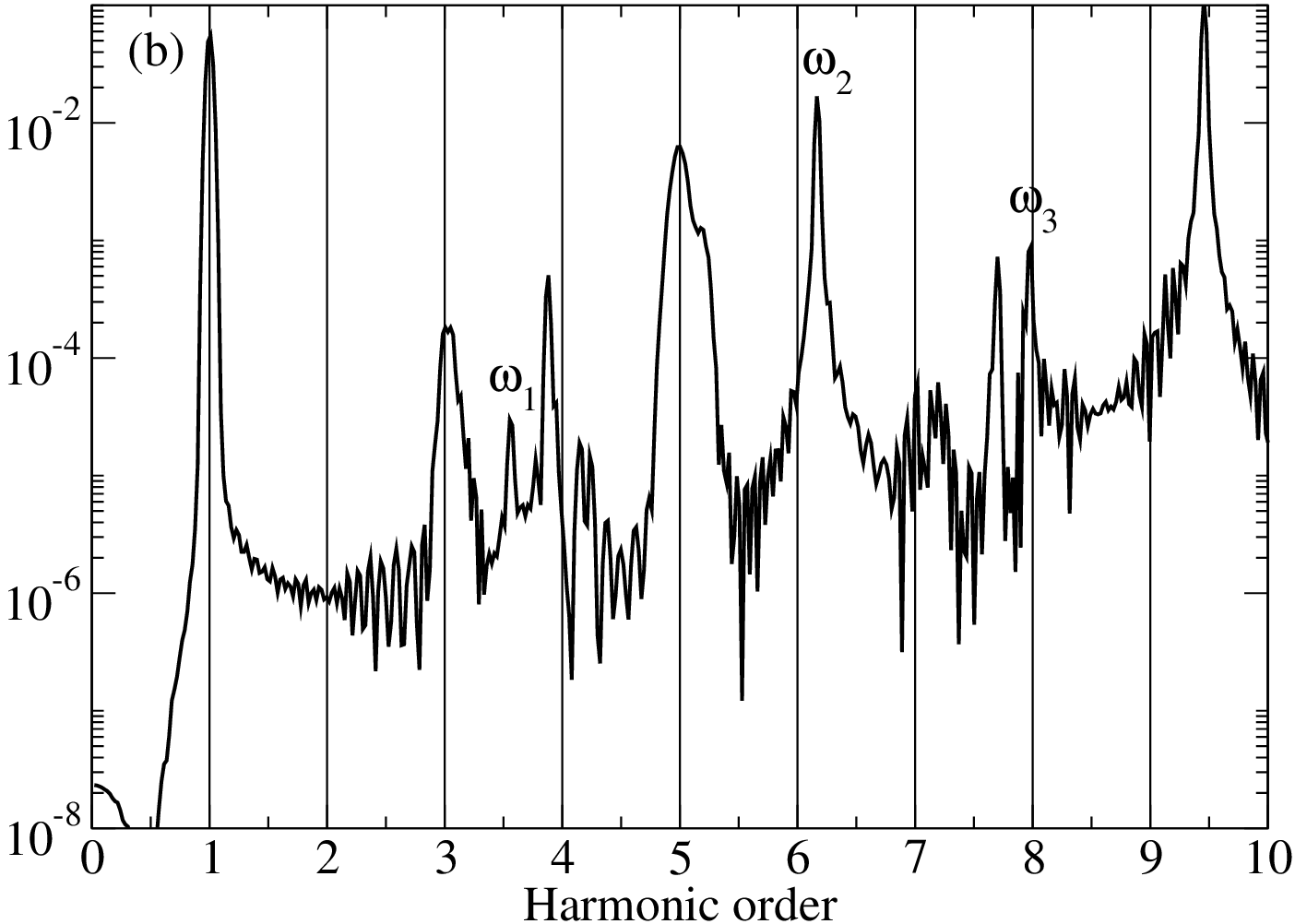}
\caption{
(a) HHG spectrum in a cavity ($L_x=L_y=L_z=2800$). The lowest cavity modes
are $\omega_1/\omega_h=1.91, \omega_2/\omega_h=3.30,
\omega_3/\omega_h=4.26$. 
(b) HHG spectrum in a cavity ($L_x=L_y=L_z=1500$). The lowest cavity modes
are $\omega_1/\omega_h=3.56, \omega_2/\omega_h=6.16,
\omega_3/\omega_h=7.96$. 
The vertical axis is $I_h$ and the horizontal axis is $\omega/\omega_h$.
}
\label{hhg3}
\end{figure}

Not only do atomic transitions appear in the HHG spectrum, but one
can find states with energies corresponding to the cavity frequency as
well. These states are coupled to matter states and appear in the
spectrum in both the absorption and HHG cases. Fig.\ref{hhg3} shows 
the peaks corresponding to the three lowest cavity 
modes for two different cavity sizes. The first (Fig.\ref{hhg3}a) is a
larger cavity with lower frequency modes. The lowest three modes
clearly show up with nearly equal intensity. Higher modes coincide
with other peaks and are not visible. The second example (Fig.\ref{hhg3}b)
shows the HHG for a smaller cavity with higher frequency modes. The 
calculations show that the appearance of peaks follows the change of
cavity frequencies. 

In the HHG calculations, we have shown the raw data without any
smoothing. Various smoothing approaches are used in HHG calculations to 
hinder the noisy oscillations and reinforce the peaks. The HHG spectrum
calculations are sensitive to the laser strength, duration, and pulse
shape. Some of the oscillations can be removed by using a much longer 
laser pulse but that is computationally expensive. The peaks
originating from the transition between coupled light-matter states 
appear with different intensities depending on the laser parameters.

\begin{figure}
\includegraphics[width=0.45\textwidth]{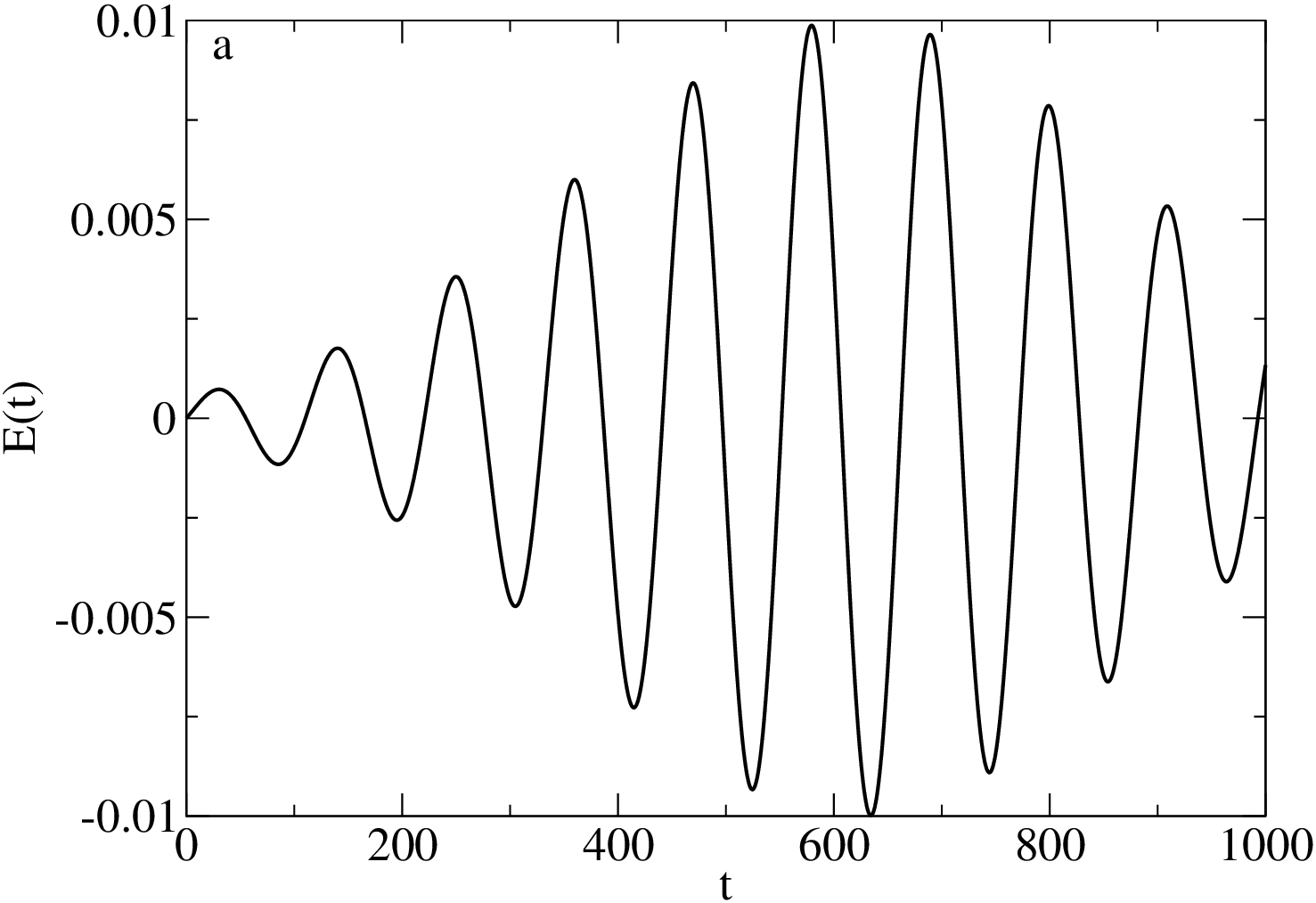}\\
\includegraphics[width=0.45\textwidth]{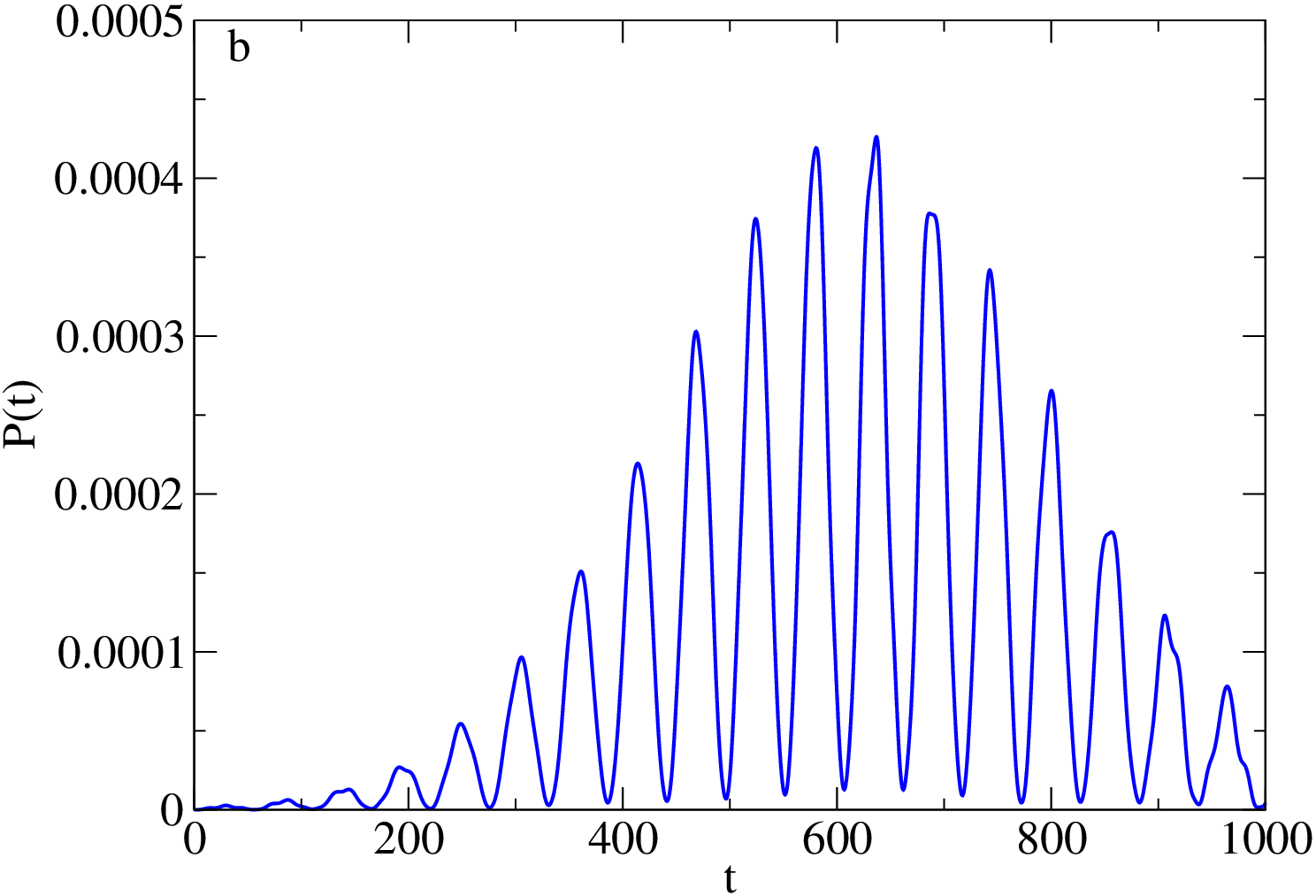}\\
\includegraphics[width=0.45\textwidth]{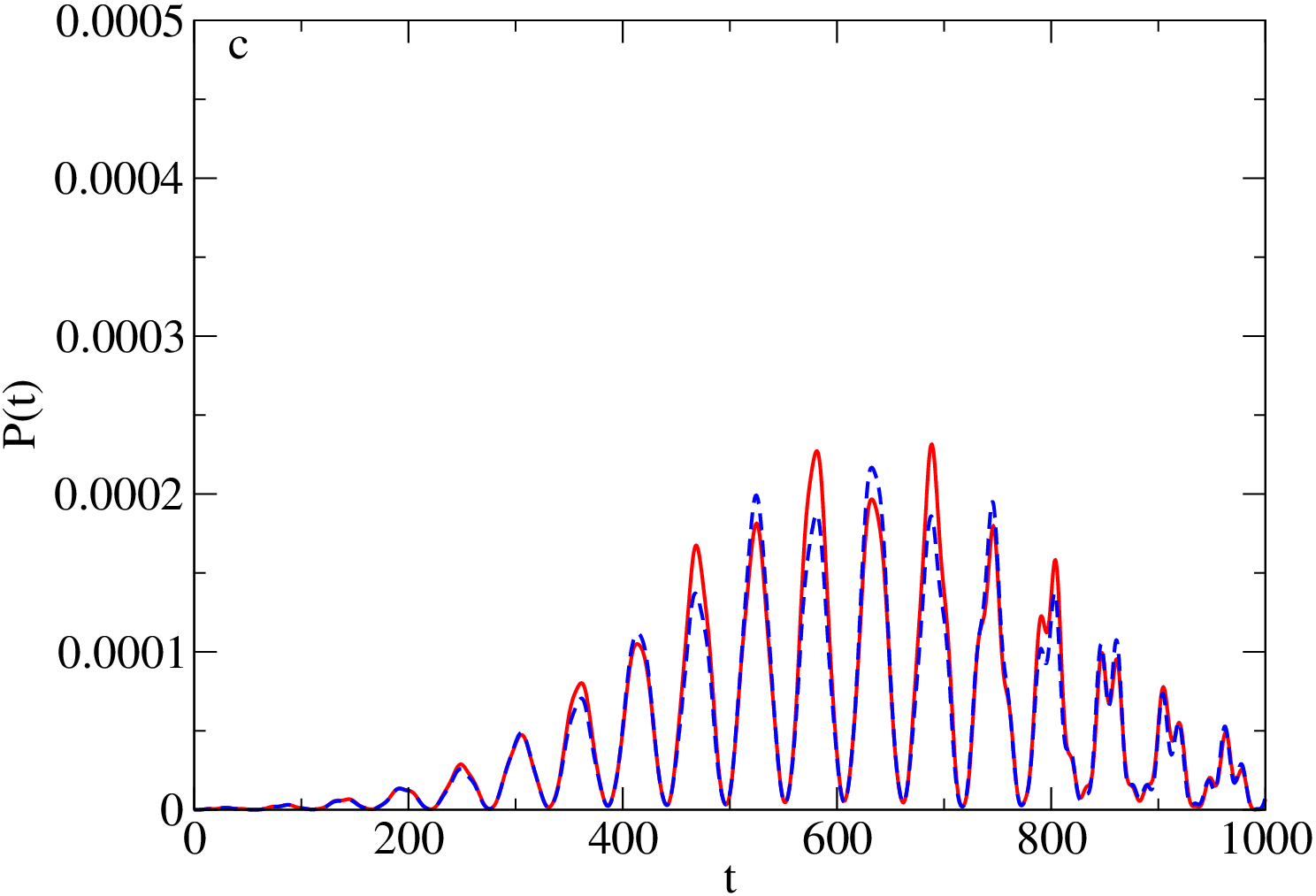}
\caption{
(a) Laser field defined as $E(t)=E_0{\rm
e}^{\frac{(t-T)^2}{\tau^2}}\sin(\omega_h t)$,
with $E_0=0.01, \omega_h=0.057 , \tau=8.8,T=15$.
(b) Evolution of the 2p state when the lowest frequency is 0.3.
(c) Evolution of the LP (solid line) and UP (dashed line) 
when the lowest frequency is 0.375. The
same cavity is used is for Fig. \ref{ale}. }
\label{pol}
\end{figure}

In Figs.\ref{ale} and \ref{ale1} we have shown how an upper polariton (UP) 
and lower polariton (LP) is
formed when the cavity frequency approaches a transition energy. 
Fig.\ref{pol} shows how a laser field (Fig.\ref{pol}a) excites
the LP and UP states. The calculation starts at $t=0$ using the ground
state wave function. In that state lowest 1s state coupled to the $\vert\mathbf{0}\rangle$ is the dominant
configuration with 0.99 probability. This states remains the dominant
configuration during the laser pulse. The laser excites other states
but the probability of those states are much smaller. A stronger laser
pulse could significantly decrease the population of the lowest state, but
our goal here is the study of the LP and UP states. 
Fig.\ref{pol}b shows
the 1s-$\omega_1$ and  the 2p-$\omega_1$ during the laser pulse. 
Both states oscillate following the laser amplitude when $\omega_1$ is below the 1s-2p
transition energy, although the probability amplitude of the
1s-$\omega_1$ state is much smaller.  
When $\omega_1$ is close to the transition frequency, the 1s-$\omega_1$ and the 2p-$\omega_1$ state 
forms the LP and UP states. These are the same states that were
studied in Fig.\ref{ale1}. The laser excites both
states similarly (Fig.\ref{pol}c) and the probability is nearly half of the non-resonant 
(Fig.\ref{pol}b,  $\omega_1=0.3$) case. This last example illustrates 
how the entangled light-matter states evolve during the laser
excitation depending on  the cavity modes. The dynamics of these
coupled light-matter states build up the HHG spectrum of the system.

\section{Summary}
A coupled light-matter basis was used to calculate the properties of one
electron systems in optical cavities. The electron's wave function was
expanded in Gaussians which provides a flexible basis to describe
excited states in laser fields. The light states are described by
using a Fock representation. The PF Hamiltonian has a term, a product of
the photon displacement operator and the matter dipole operator, that
couples the light and matter. This term is conveniently handled by out
tensor product basis of light and matter states. 

We have shown how the light-matter states change as a function of the
cavity size. The cavity dimension determines the cavity frequencies
and the mode functions, $\omega_i$ and $\lambda_i$. In a small cavity 
$\lambda_i$ is larger but $\omega_i$ also increases and it becomes much
larger than the lowest transition states of the coupled system. This
makes the manipulation of the resonances near transition energies more
difficult. The coupled states are raising with increasing $\omega_i$
leading the a myriad of avoided level crossings (Rabi splittings). The
example of 1s,2s, and 2p states shows that several energy level can
repel each other at a single point due to the coupling to light.

The avoided level crossing appears as a Rabi splitting in the absorption
spectrum. Due to the repulsion of the levels, there are no available
states to absorb light in the energy gap leading to electromagnetically 
induced transparency. We have shown that any frequency mode in the
cavity can be fine-tuned to split any absorption peak. 

The effect of a cavity on the HHG spectrum was also studied. Strong
coupling breaks the symmetry of the system and even harmonics can
also appear. The energy spectrum of the light-matter coupled system 
is also imprinted in the HHG spectrum. This also means that there are HHG
peaks at cavity frequencies which might be used to fine tune the emitted
radiation. Besides the HHG peak at the integer multiples of the
frequency of the exciting laser, one can create peaks at desired
frequencies. The mark of the HHG spectrum can be found in all photon
spaces, each photon carries information about the electronic levels of the
emitter and the frequency of other photons as well. 

The present approach was concentrated on a simple system where
analytical matrix elements were available. This approach can also be
used for the QED-TDDFT formed by tensor coupled real space and Fock
space\cite{10.1063/5.0123909} extending the approach for multiple 
photon modes. Such calculations are under development in our group.

\begin{acknowledgments} 
This work has been supported by the National Science Foundation (NSF)
under Grant No. 2217759.
\end{acknowledgments}

\begin{thebibliography}{56}%
\makeatletter
\providecommand \@ifxundefined [1]{%
 \@ifx{#1\undefined}
}%
\providecommand \@ifnum [1]{%
 \ifnum #1\expandafter \@firstoftwo
 \else \expandafter \@secondoftwo
 \fi
}%
\providecommand \@ifx [1]{%
 \ifx #1\expandafter \@firstoftwo
 \else \expandafter \@secondoftwo
 \fi
}%
\providecommand \natexlab [1]{#1}%
\providecommand \enquote  [1]{``#1''}%
\providecommand \bibnamefont  [1]{#1}%
\providecommand \bibfnamefont [1]{#1}%
\providecommand \citenamefont [1]{#1}%
\providecommand \href@noop [0]{\@secondoftwo}%
\providecommand \href [0]{\begingroup \@sanitize@url \@href}%
\providecommand \@href[1]{\@@startlink{#1}\@@href}%
\providecommand \@@href[1]{\endgroup#1\@@endlink}%
\providecommand \@sanitize@url [0]{\catcode `\\12\catcode `\$12\catcode
  `\&12\catcode `\#12\catcode `\^12\catcode `\_12\catcode `\%12\relax}%
\providecommand \@@startlink[1]{}%
\providecommand \@@endlink[0]{}%
\providecommand \url  [0]{\begingroup\@sanitize@url \@url }%
\providecommand \@url [1]{\endgroup\@href {#1}{\urlprefix }}%
\providecommand \urlprefix  [0]{URL }%
\providecommand \Eprint [0]{\href }%
\providecommand \doibase [0]{https://doi.org/}%
\providecommand \selectlanguage [0]{\@gobble}%
\providecommand \bibinfo  [0]{\@secondoftwo}%
\providecommand \bibfield  [0]{\@secondoftwo}%
\providecommand \translation [1]{[#1]}%
\providecommand \BibitemOpen [0]{}%
\providecommand \bibitemStop [0]{}%
\providecommand \bibitemNoStop [0]{.\EOS\space}%
\providecommand \EOS [0]{\spacefactor3000\relax}%
\providecommand \BibitemShut  [1]{\csname bibitem#1\endcsname}%
\let\auto@bib@innerbib\@empty
\bibitem [{\citenamefont {Aupiais}\ \emph {et~al.}(2023)\citenamefont
  {Aupiais}, \citenamefont {Grasset}, \citenamefont {Guo}, \citenamefont
  {Daineka}, \citenamefont {Briatico}, \citenamefont {Houver}, \citenamefont
  {Perfetti}, \citenamefont {Hugonin}, \citenamefont {Greffet},\ and\
  \citenamefont {Laplace}}]{Aupiais2023}%
  \BibitemOpen
  \bibfield  {author} {\bibinfo {author} {\bibfnamefont {I.}~\bibnamefont
  {Aupiais}}, \bibinfo {author} {\bibfnamefont {R.}~\bibnamefont {Grasset}},
  \bibinfo {author} {\bibfnamefont {T.}~\bibnamefont {Guo}}, \bibinfo {author}
  {\bibfnamefont {D.}~\bibnamefont {Daineka}}, \bibinfo {author} {\bibfnamefont
  {J.}~\bibnamefont {Briatico}}, \bibinfo {author} {\bibfnamefont
  {S.}~\bibnamefont {Houver}}, \bibinfo {author} {\bibfnamefont
  {L.}~\bibnamefont {Perfetti}}, \bibinfo {author} {\bibfnamefont {J.-P.}\
  \bibnamefont {Hugonin}}, \bibinfo {author} {\bibfnamefont {J.-J.}\
  \bibnamefont {Greffet}},\ and\ \bibinfo {author} {\bibfnamefont
  {Y.}~\bibnamefont {Laplace}},\ }\bibfield  {title} {\bibinfo {title}
  {Ultrasmall and tunable terahertz surface plasmon cavities at the ultimate
  plasmonic limit},\ }\href {https://doi.org/10.1038/s41467-023-43394-w}
  {\bibfield  {journal} {\bibinfo  {journal} {Nature Communications}\ }\textbf
  {\bibinfo {volume} {14}},\ \bibinfo {pages} {7645} (\bibinfo {year}
  {2023})}\BibitemShut {NoStop}%
\bibitem [{\citenamefont {Baumberg}(2022)}]{doi:10.1021/acs.nanolett.2c01695}%
  \BibitemOpen
  \bibfield  {author} {\bibinfo {author} {\bibfnamefont {J.~J.}\ \bibnamefont
  {Baumberg}},\ }\bibfield  {title} {\bibinfo {title} {Picocavities: a
  primer},\ }\href {https://doi.org/10.1021/acs.nanolett.2c01695} {\bibfield
  {journal} {\bibinfo  {journal} {Nano Letters}\ }\textbf {\bibinfo {volume}
  {22}},\ \bibinfo {pages} {5859} (\bibinfo {year} {2022})}\BibitemShut
  {NoStop}%
\bibitem [{\citenamefont {Bhattacharya}\ \emph {et~al.}(2023)\citenamefont
  {Bhattacharya}, \citenamefont {Lamprou}, \citenamefont {Maxwell},
  \citenamefont {Ordóñez}, \citenamefont {Pisanty}, \citenamefont
  {Rivera-Dean}, \citenamefont {Stammer}, \citenamefont {Ciappina},
  \citenamefont {Lewenstein},\ and\ \citenamefont
  {Tzallas}}]{Bhattacharya_2023}%
  \BibitemOpen
  \bibfield  {author} {\bibinfo {author} {\bibfnamefont {U.}~\bibnamefont
  {Bhattacharya}}, \bibinfo {author} {\bibfnamefont {T.}~\bibnamefont
  {Lamprou}}, \bibinfo {author} {\bibfnamefont {A.~S.}\ \bibnamefont
  {Maxwell}}, \bibinfo {author} {\bibfnamefont {A.}~\bibnamefont {Ordóñez}},
  \bibinfo {author} {\bibfnamefont {E.}~\bibnamefont {Pisanty}}, \bibinfo
  {author} {\bibfnamefont {J.}~\bibnamefont {Rivera-Dean}}, \bibinfo {author}
  {\bibfnamefont {P.}~\bibnamefont {Stammer}}, \bibinfo {author} {\bibfnamefont
  {M.~F.}\ \bibnamefont {Ciappina}}, \bibinfo {author} {\bibfnamefont
  {M.}~\bibnamefont {Lewenstein}},\ and\ \bibinfo {author} {\bibfnamefont
  {P.}~\bibnamefont {Tzallas}},\ }\bibfield  {title} {\bibinfo {title}
  {Strong–laser–field physics, non–classical light states and quantum
  information science},\ }\href {https://doi.org/10.1088/1361-6633/acea31}
  {\bibfield  {journal} {\bibinfo  {journal} {Reports on Progress in Physics}\
  }\textbf {\bibinfo {volume} {86}},\ \bibinfo {pages} {094401} (\bibinfo
  {year} {2023})}\BibitemShut {NoStop}%
\bibitem [{\citenamefont {Gorlach}\ \emph {et~al.}(2023)\citenamefont
  {Gorlach}, \citenamefont {Tzur}, \citenamefont {Birk}, \citenamefont
  {Kruger}, \citenamefont {Rivera}, \citenamefont {Cohen},\ and\ \citenamefont
  {Kaminer}}]{Gorlach2023}%
  \BibitemOpen
  \bibfield  {author} {\bibinfo {author} {\bibfnamefont {A.}~\bibnamefont
  {Gorlach}}, \bibinfo {author} {\bibfnamefont {M.~E.}\ \bibnamefont {Tzur}},
  \bibinfo {author} {\bibfnamefont {M.}~\bibnamefont {Birk}}, \bibinfo {author}
  {\bibfnamefont {M.}~\bibnamefont {Kruger}}, \bibinfo {author} {\bibfnamefont
  {N.}~\bibnamefont {Rivera}}, \bibinfo {author} {\bibfnamefont
  {O.}~\bibnamefont {Cohen}},\ and\ \bibinfo {author} {\bibfnamefont
  {I.}~\bibnamefont {Kaminer}},\ }\bibfield  {title} {\bibinfo {title}
  {High-harmonic generation driven by quantum light},\ }\href
  {https://doi.org/10.1038/s41567-023-02127-y} {\bibfield  {journal} {\bibinfo
  {journal} {Nature Physics}\ }\textbf {\bibinfo {volume} {19}},\ \bibinfo
  {pages} {1689} (\bibinfo {year} {2023})}\BibitemShut {NoStop}%
\bibitem [{\citenamefont {Bogatskaya}\ \emph
  {et~al.}(2017{\natexlab{a}})\citenamefont {Bogatskaya}, \citenamefont
  {Volkova},\ and\ \citenamefont {Popov}}]{Bogatskaya2017}%
  \BibitemOpen
  \bibfield  {author} {\bibinfo {author} {\bibfnamefont {A.~V.}\ \bibnamefont
  {Bogatskaya}}, \bibinfo {author} {\bibfnamefont {E.~A.}\ \bibnamefont
  {Volkova}},\ and\ \bibinfo {author} {\bibfnamefont {A.~M.}\ \bibnamefont
  {Popov}},\ }\bibfield  {title} {\bibinfo {title} {Spontaneous emission of
  atoms in a strong laser field},\ }\href
  {https://doi.org/10.1134/S1063776117090114} {\bibfield  {journal} {\bibinfo
  {journal} {Journal of Experimental and Theoretical Physics}\ }\textbf
  {\bibinfo {volume} {125}},\ \bibinfo {pages} {587} (\bibinfo {year}
  {2017}{\natexlab{a}})}\BibitemShut {NoStop}%
\bibitem [{\citenamefont {Wang}\ and\ \citenamefont
  {Lai}(2023)}]{PhysRevA.108.063101}%
  \BibitemOpen
  \bibfield  {author} {\bibinfo {author} {\bibfnamefont {S.}~\bibnamefont
  {Wang}}\ and\ \bibinfo {author} {\bibfnamefont {X.}~\bibnamefont {Lai}},\
  }\bibfield  {title} {\bibinfo {title} {High-order above-threshold ionization
  of an atom in intense quantum light},\ }\href
  {https://doi.org/10.1103/PhysRevA.108.063101} {\bibfield  {journal} {\bibinfo
   {journal} {Phys. Rev. A}\ }\textbf {\bibinfo {volume} {108}},\ \bibinfo
  {pages} {063101} (\bibinfo {year} {2023})}\BibitemShut {NoStop}%
\bibitem [{\citenamefont {Rivera-Dean}\ \emph {et~al.}(2024)\citenamefont
  {Rivera-Dean}, \citenamefont {Stammer}, \citenamefont {Maxwell},
  \citenamefont {Lamprou}, \citenamefont {Pisanty}, \citenamefont {Tzallas},
  \citenamefont {Lewenstein},\ and\ \citenamefont
  {Ciappina}}]{PhysRevA.109.033706}%
  \BibitemOpen
  \bibfield  {author} {\bibinfo {author} {\bibfnamefont {J.}~\bibnamefont
  {Rivera-Dean}}, \bibinfo {author} {\bibfnamefont {P.}~\bibnamefont
  {Stammer}}, \bibinfo {author} {\bibfnamefont {A.~S.}\ \bibnamefont
  {Maxwell}}, \bibinfo {author} {\bibfnamefont {T.}~\bibnamefont {Lamprou}},
  \bibinfo {author} {\bibfnamefont {E.}~\bibnamefont {Pisanty}}, \bibinfo
  {author} {\bibfnamefont {P.}~\bibnamefont {Tzallas}}, \bibinfo {author}
  {\bibfnamefont {M.}~\bibnamefont {Lewenstein}},\ and\ \bibinfo {author}
  {\bibfnamefont {M.~F.}\ \bibnamefont {Ciappina}},\ }\bibfield  {title}
  {\bibinfo {title} {Quantum-optical analysis of high-order harmonic generation
  in h$_2^+$ molecules},\ }\href {https://doi.org/10.1103/PhysRevA.109.033706}
  {\bibfield  {journal} {\bibinfo  {journal} {Phys. Rev. A}\ }\textbf {\bibinfo
  {volume} {109}},\ \bibinfo {pages} {033706} (\bibinfo {year}
  {2024})}\BibitemShut {NoStop}%
\bibitem [{\citenamefont {Moiseyev}\ and\ \citenamefont
  {Tzur}(2023)}]{Moiseyev_2024}%
  \BibitemOpen
  \bibfield  {author} {\bibinfo {author} {\bibfnamefont {N.}~\bibnamefont
  {Moiseyev}}\ and\ \bibinfo {author} {\bibfnamefont {M.~E.}\ \bibnamefont
  {Tzur}},\ }\bibfield  {title} {\bibinfo {title} {The conditions for the
  analog of qed photons in semi-classical periodically driven systems},\ }\href
  {https://doi.org/10.1088/2040-8986/ad15eb} {\bibfield  {journal} {\bibinfo
  {journal} {Journal of Optics}\ }\textbf {\bibinfo {volume} {26}},\ \bibinfo
  {pages} {025501} (\bibinfo {year} {2023})}\BibitemShut {NoStop}%
\bibitem [{\citenamefont {Bogatskaya}\ and\ \citenamefont
  {Popov}(2020)}]{Bogatskaya_2020}%
  \BibitemOpen
  \bibfield  {author} {\bibinfo {author} {\bibfnamefont {A.~V.}\ \bibnamefont
  {Bogatskaya}}\ and\ \bibinfo {author} {\bibfnamefont {A.~M.}\ \bibnamefont
  {Popov}},\ }\bibfield  {title} {\bibinfo {title} {Limitations of the
  semiclassical approach for the problem of two-level atomic ensemble emission
  in a resonant laser field},\ }\href
  {https://doi.org/10.1088/1612-202X/aba198} {\bibfield  {journal} {\bibinfo
  {journal} {Laser Physics Letters}\ }\textbf {\bibinfo {volume} {17}},\
  \bibinfo {pages} {096002} (\bibinfo {year} {2020})}\BibitemShut {NoStop}%
\bibitem [{\citenamefont {Peng}\ \emph {et~al.}(2023)\citenamefont {Peng},
  \citenamefont {Hu}, \citenamefont {Liu}, \citenamefont {Liu}, \citenamefont
  {Zhao},\ and\ \citenamefont {Yuan}}]{PhysRevA.108.053119}%
  \BibitemOpen
  \bibfield  {author} {\bibinfo {author} {\bibfnamefont {Z.}~\bibnamefont
  {Peng}}, \bibinfo {author} {\bibfnamefont {H.}~\bibnamefont {Hu}}, \bibinfo
  {author} {\bibfnamefont {J.}~\bibnamefont {Liu}}, \bibinfo {author}
  {\bibfnamefont {J.}~\bibnamefont {Liu}}, \bibinfo {author} {\bibfnamefont
  {Z.}~\bibnamefont {Zhao}},\ and\ \bibinfo {author} {\bibfnamefont
  {J.}~\bibnamefont {Yuan}},\ }\bibfield  {title} {\bibinfo {title} {Quantum
  radiation and absorption by current fluctuations of atoms in strong laser
  fields},\ }\href {https://doi.org/10.1103/PhysRevA.108.053119} {\bibfield
  {journal} {\bibinfo  {journal} {Phys. Rev. A}\ }\textbf {\bibinfo {volume}
  {108}},\ \bibinfo {pages} {053119} (\bibinfo {year} {2023})}\BibitemShut
  {NoStop}%
\bibitem [{\citenamefont {Ebbesen}(2016)}]{doi:10.1021/acs.accounts.6b00295}%
  \BibitemOpen
  \bibfield  {author} {\bibinfo {author} {\bibfnamefont {T.~W.}\ \bibnamefont
  {Ebbesen}},\ }\bibfield  {title} {\bibinfo {title} {Hybrid light–matter
  states in a molecular and material science perspective},\ }\href
  {https://doi.org/10.1021/acs.accounts.6b00295} {\bibfield  {journal}
  {\bibinfo  {journal} {Accounts of Chemical Research}\ }\textbf {\bibinfo
  {volume} {49}},\ \bibinfo {pages} {2403} (\bibinfo {year}
  {2016})}\BibitemShut {NoStop}%
\bibitem [{\citenamefont {Li}\ \emph {et~al.}(2022)\citenamefont {Li},
  \citenamefont {Cui}, \citenamefont {Subotnik},\ and\ \citenamefont
  {Nitzan}}]{doi:10.1146/annurev-physchem-090519-042621}%
  \BibitemOpen
  \bibfield  {author} {\bibinfo {author} {\bibfnamefont {T.~E.}\ \bibnamefont
  {Li}}, \bibinfo {author} {\bibfnamefont {B.}~\bibnamefont {Cui}}, \bibinfo
  {author} {\bibfnamefont {J.~E.}\ \bibnamefont {Subotnik}},\ and\ \bibinfo
  {author} {\bibfnamefont {A.}~\bibnamefont {Nitzan}},\ }\bibfield  {title}
  {\bibinfo {title} {Molecular polaritonics: Chemical dynamics under strong
  light–matter coupling},\ }\href
  {https://doi.org/10.1146/annurev-physchem-090519-042621} {\bibfield
  {journal} {\bibinfo  {journal} {Annual Review of Physical Chemistry}\
  }\textbf {\bibinfo {volume} {73}},\ \bibinfo {pages} {43} (\bibinfo {year}
  {2022})}\BibitemShut {NoStop}%
\bibitem [{\citenamefont {Ruggenthaler}\ \emph {et~al.}(2018)\citenamefont
  {Ruggenthaler}, \citenamefont {Tancogne-Dejean}, \citenamefont {Flick},
  \citenamefont {Appel},\ and\ \citenamefont {Rubio}}]{Ruggenthaler2018}%
  \BibitemOpen
  \bibfield  {author} {\bibinfo {author} {\bibfnamefont {M.}~\bibnamefont
  {Ruggenthaler}}, \bibinfo {author} {\bibfnamefont {N.}~\bibnamefont
  {Tancogne-Dejean}}, \bibinfo {author} {\bibfnamefont {J.}~\bibnamefont
  {Flick}}, \bibinfo {author} {\bibfnamefont {H.}~\bibnamefont {Appel}},\ and\
  \bibinfo {author} {\bibfnamefont {A.}~\bibnamefont {Rubio}},\ }\bibfield
  {title} {\bibinfo {title} {From a quantum-electrodynamical light--matter
  description to novel spectroscopies},\ }\href
  {https://doi.org/10.1038/s41570-018-0118} {\bibfield  {journal} {\bibinfo
  {journal} {Nature Reviews Chemistry}\ }\textbf {\bibinfo {volume} {2}},\
  \bibinfo {pages} {0118} (\bibinfo {year} {2018})}\BibitemShut {NoStop}%
\bibitem [{\citenamefont {McTague}\ and\ \citenamefont
  {Foley}(2022)}]{mctague}%
  \BibitemOpen
  \bibfield  {author} {\bibinfo {author} {\bibfnamefont {J.}~\bibnamefont
  {McTague}}\ and\ \bibinfo {author} {\bibfnamefont {J.~J.}\ \bibnamefont
  {Foley}},\ }\bibfield  {title} {\bibinfo {title} {Non-hermitian cavity
  quantum electrodynamics–configuration interaction singles approach for
  polaritonic structure with ab initio molecular hamiltonians},\ }\href
  {https://doi.org/10.1063/5.0091953} {\bibfield  {journal} {\bibinfo
  {journal} {The Journal of Chemical Physics}\ }\textbf {\bibinfo {volume}
  {156}},\ \bibinfo {pages} {154103} (\bibinfo {year} {2022})}\BibitemShut
  {NoStop}%
\bibitem [{\citenamefont {Sánchez-Barquilla}\ \emph
  {et~al.}(2022)\citenamefont {Sánchez-Barquilla}, \citenamefont
  {Fernández-Domínguez}, \citenamefont {Feist},\ and\ \citenamefont
  {García-Vidal}}]{doi:10.1021/acsphotonics.2c00048}%
  \BibitemOpen
  \bibfield  {author} {\bibinfo {author} {\bibfnamefont {M.}~\bibnamefont
  {Sánchez-Barquilla}}, \bibinfo {author} {\bibfnamefont {A.~I.}\ \bibnamefont
  {Fernández-Domínguez}}, \bibinfo {author} {\bibfnamefont {J.}~\bibnamefont
  {Feist}},\ and\ \bibinfo {author} {\bibfnamefont {F.~J.}\ \bibnamefont
  {García-Vidal}},\ }\bibfield  {title} {\bibinfo {title} {A theoretical
  perspective on molecular polaritonics},\ }\href
  {https://doi.org/10.1021/acsphotonics.2c00048} {\bibfield  {journal}
  {\bibinfo  {journal} {ACS Photonics}\ }\textbf {\bibinfo {volume} {9}},\
  \bibinfo {pages} {1830} (\bibinfo {year} {2022})}\BibitemShut {NoStop}%
\bibitem [{\citenamefont {Mallory}\ and\ \citenamefont {DePrince}(2022)}]{rd}%
  \BibitemOpen
  \bibfield  {author} {\bibinfo {author} {\bibfnamefont {J.~D.}\ \bibnamefont
  {Mallory}}\ and\ \bibinfo {author} {\bibfnamefont {A.~E.}\ \bibnamefont
  {DePrince}},\ }\bibfield  {title} {\bibinfo {title}
  {Reduced-density-matrix-based ab initio cavity quantum electrodynamics},\
  }\href {https://doi.org/10.1103/PhysRevA.106.053710} {\bibfield  {journal}
  {\bibinfo  {journal} {Phys. Rev. A}\ }\textbf {\bibinfo {volume} {106}},\
  \bibinfo {pages} {053710} (\bibinfo {year} {2022})}\BibitemShut {NoStop}%
\bibitem [{\citenamefont {Ahrens}\ \emph {et~al.}(2021)\citenamefont {Ahrens},
  \citenamefont {Huang}, \citenamefont {Beutel}, \citenamefont {Covington},\
  and\ \citenamefont {Varga}}]{PhysRevLett.127.273601}%
  \BibitemOpen
  \bibfield  {author} {\bibinfo {author} {\bibfnamefont {A.}~\bibnamefont
  {Ahrens}}, \bibinfo {author} {\bibfnamefont {C.}~\bibnamefont {Huang}},
  \bibinfo {author} {\bibfnamefont {M.}~\bibnamefont {Beutel}}, \bibinfo
  {author} {\bibfnamefont {C.}~\bibnamefont {Covington}},\ and\ \bibinfo
  {author} {\bibfnamefont {K.}~\bibnamefont {Varga}},\ }\bibfield  {title}
  {\bibinfo {title} {Stochastic variational approach to small atoms and
  molecules coupled to quantum field modes in cavity qed},\ }\href
  {https://doi.org/10.1103/PhysRevLett.127.273601} {\bibfield  {journal}
  {\bibinfo  {journal} {Phys. Rev. Lett.}\ }\textbf {\bibinfo {volume} {127}},\
  \bibinfo {pages} {273601} (\bibinfo {year} {2021})}\BibitemShut {NoStop}%
\bibitem [{\citenamefont {Sch\"afer}\ and\ \citenamefont
  {Johansson}(2022)}]{PhysRevLett.128.156402}%
  \BibitemOpen
  \bibfield  {author} {\bibinfo {author} {\bibfnamefont {C.}~\bibnamefont
  {Sch\"afer}}\ and\ \bibinfo {author} {\bibfnamefont {G.}~\bibnamefont
  {Johansson}},\ }\bibfield  {title} {\bibinfo {title} {Shortcut to
  self-consistent light-matter interaction and realistic spectra from first
  principles},\ }\href {https://doi.org/10.1103/PhysRevLett.128.156402}
  {\bibfield  {journal} {\bibinfo  {journal} {Phys. Rev. Lett.}\ }\textbf
  {\bibinfo {volume} {128}},\ \bibinfo {pages} {156402} (\bibinfo {year}
  {2022})}\BibitemShut {NoStop}%
\bibitem [{\citenamefont {Malave}\ \emph
  {et~al.}(2022{\natexlab{a}})\citenamefont {Malave}, \citenamefont {Ahrens},
  \citenamefont {Pitagora}, \citenamefont {Covington},\ and\ \citenamefont
  {Varga}}]{10.1063/5.0123909}%
  \BibitemOpen
  \bibfield  {author} {\bibinfo {author} {\bibfnamefont {J.}~\bibnamefont
  {Malave}}, \bibinfo {author} {\bibfnamefont {A.}~\bibnamefont {Ahrens}},
  \bibinfo {author} {\bibfnamefont {D.}~\bibnamefont {Pitagora}}, \bibinfo
  {author} {\bibfnamefont {C.}~\bibnamefont {Covington}},\ and\ \bibinfo
  {author} {\bibfnamefont {K.}~\bibnamefont {Varga}},\ }\bibfield  {title}
  {\bibinfo {title} {{Real-space, real-time approach to
  quantum-electrodynamical time-dependent density functional theory}},\ }\href
  {https://doi.org/10.1063/5.0123909} {\bibfield  {journal} {\bibinfo
  {journal} {The Journal of Chemical Physics}\ }\textbf {\bibinfo {volume}
  {157}},\ \bibinfo {pages} {194106} (\bibinfo {year}
  {2022}{\natexlab{a}})}\BibitemShut {NoStop}%
\bibitem [{\citenamefont {Huang}\ \emph {et~al.}(2023)\citenamefont {Huang},
  \citenamefont {Covington},\ and\ \citenamefont
  {Varga}}]{PhysRevB.107.235130}%
  \BibitemOpen
  \bibfield  {author} {\bibinfo {author} {\bibfnamefont {C.}~\bibnamefont
  {Huang}}, \bibinfo {author} {\bibfnamefont {C.}~\bibnamefont {Covington}},\
  and\ \bibinfo {author} {\bibfnamefont {K.}~\bibnamefont {Varga}},\ }\bibfield
   {title} {\bibinfo {title} {Harmonically confined $n$-electron systems
  coupled to light in a cavity: Time-dependent case},\ }\href
  {https://doi.org/10.1103/PhysRevB.107.235130} {\bibfield  {journal} {\bibinfo
   {journal} {Phys. Rev. B}\ }\textbf {\bibinfo {volume} {107}},\ \bibinfo
  {pages} {235130} (\bibinfo {year} {2023})}\BibitemShut {NoStop}%
\bibitem [{\citenamefont {Malave}\ \emph
  {et~al.}(2022{\natexlab{b}})\citenamefont {Malave}, \citenamefont {Aklilu},
  \citenamefont {Beutel}, \citenamefont {Huang},\ and\ \citenamefont
  {Varga}}]{PhysRevB.105.115127}%
  \BibitemOpen
  \bibfield  {author} {\bibinfo {author} {\bibfnamefont {J.}~\bibnamefont
  {Malave}}, \bibinfo {author} {\bibfnamefont {Y.~S.}\ \bibnamefont {Aklilu}},
  \bibinfo {author} {\bibfnamefont {M.}~\bibnamefont {Beutel}}, \bibinfo
  {author} {\bibfnamefont {C.}~\bibnamefont {Huang}},\ and\ \bibinfo {author}
  {\bibfnamefont {K.}~\bibnamefont {Varga}},\ }\bibfield  {title} {\bibinfo
  {title} {Harmonically confined $n$-electron systems coupled to light in a
  cavity},\ }\href {https://doi.org/10.1103/PhysRevB.105.115127} {\bibfield
  {journal} {\bibinfo  {journal} {Phys. Rev. B}\ }\textbf {\bibinfo {volume}
  {105}},\ \bibinfo {pages} {115127} (\bibinfo {year}
  {2022}{\natexlab{b}})}\BibitemShut {NoStop}%
\bibitem [{\citenamefont {Huang}\ \emph {et~al.}(2021)\citenamefont {Huang},
  \citenamefont {Ahrens}, \citenamefont {Beutel},\ and\ \citenamefont
  {Varga}}]{PhysRevB.104.165147}%
  \BibitemOpen
  \bibfield  {author} {\bibinfo {author} {\bibfnamefont {C.}~\bibnamefont
  {Huang}}, \bibinfo {author} {\bibfnamefont {A.}~\bibnamefont {Ahrens}},
  \bibinfo {author} {\bibfnamefont {M.}~\bibnamefont {Beutel}},\ and\ \bibinfo
  {author} {\bibfnamefont {K.}~\bibnamefont {Varga}},\ }\bibfield  {title}
  {\bibinfo {title} {Two electrons in harmonic confinement coupled to light in
  a cavity},\ }\href {https://doi.org/10.1103/PhysRevB.104.165147} {\bibfield
  {journal} {\bibinfo  {journal} {Phys. Rev. B}\ }\textbf {\bibinfo {volume}
  {104}},\ \bibinfo {pages} {165147} (\bibinfo {year} {2021})}\BibitemShut
  {NoStop}%
\bibitem [{\citenamefont {Bogatskaya}\ \emph
  {et~al.}(2017{\natexlab{b}})\citenamefont {Bogatskaya}, \citenamefont
  {Volkova},\ and\ \citenamefont {Popov}}]{Bogatskaya_2017}%
  \BibitemOpen
  \bibfield  {author} {\bibinfo {author} {\bibfnamefont {A.~V.}\ \bibnamefont
  {Bogatskaya}}, \bibinfo {author} {\bibfnamefont {E.~A.}\ \bibnamefont
  {Volkova}},\ and\ \bibinfo {author} {\bibfnamefont {A.~M.}\ \bibnamefont
  {Popov}},\ }\bibfield  {title} {\bibinfo {title} {Spontaneous emission from
  the atom stabilized by a strong high-frequency laser field},\ }\href
  {https://doi.org/10.1088/1555-6611/aa7f9f} {\bibfield  {journal} {\bibinfo
  {journal} {Laser Physics}\ }\textbf {\bibinfo {volume} {27}},\ \bibinfo
  {pages} {095302} (\bibinfo {year} {2017}{\natexlab{b}})}\BibitemShut
  {NoStop}%
\bibitem [{\citenamefont {Gombk\"ot\ifmmode~\mbox{\H{o}}\else \H{o}\fi{}}\
  \emph {et~al.}(2020)\citenamefont {Gombk\"ot\ifmmode~\mbox{\H{o}}\else
  \H{o}\fi{}}, \citenamefont {Varr\'o}, \citenamefont {Mati},\ and\
  \citenamefont {F\"oldi}}]{PhysRevA.101.013418}%
  \BibitemOpen
  \bibfield  {author} {\bibinfo {author} {\bibfnamefont {A.}~\bibnamefont
  {Gombk\"ot\ifmmode~\mbox{\H{o}}\else \H{o}\fi{}}}, \bibinfo {author}
  {\bibfnamefont {S.}~\bibnamefont {Varr\'o}}, \bibinfo {author} {\bibfnamefont
  {P.}~\bibnamefont {Mati}},\ and\ \bibinfo {author} {\bibfnamefont
  {P.}~\bibnamefont {F\"oldi}},\ }\bibfield  {title} {\bibinfo {title}
  {High-order harmonic generation as induced by a quantized field: Phase-space
  picture},\ }\href {https://doi.org/10.1103/PhysRevA.101.013418} {\bibfield
  {journal} {\bibinfo  {journal} {Phys. Rev. A}\ }\textbf {\bibinfo {volume}
  {101}},\ \bibinfo {pages} {013418} (\bibinfo {year} {2020})}\BibitemShut
  {NoStop}%
\bibitem [{\citenamefont {Varró}(2021)}]{photonics8070269}%
  \BibitemOpen
  \bibfield  {author} {\bibinfo {author} {\bibfnamefont {S.}~\bibnamefont
  {Varró}},\ }\bibfield  {title} {\bibinfo {title} {Quantum optical aspects of
  high-harmonic generation},\ }\bibfield  {journal} {\bibinfo  {journal}
  {Photonics}\ }\textbf {\bibinfo {volume} {8}},\ \href
  {https://doi.org/10.3390/photonics8070269} {10.3390/photonics8070269}
  (\bibinfo {year} {2021})\BibitemShut {NoStop}%
\bibitem [{\citenamefont {Gombk\"ot\ifmmode~\mbox{\H{o}}\else \H{o}\fi{}}\
  \emph {et~al.}(2021)\citenamefont {Gombk\"ot\ifmmode~\mbox{\H{o}}\else
  \H{o}\fi{}}, \citenamefont {F\"oldi},\ and\ \citenamefont
  {Varr\'o}}]{PhysRevA.104.033703}%
  \BibitemOpen
  \bibfield  {author} {\bibinfo {author} {\bibfnamefont {A.}~\bibnamefont
  {Gombk\"ot\ifmmode~\mbox{\H{o}}\else \H{o}\fi{}}}, \bibinfo {author}
  {\bibfnamefont {P.}~\bibnamefont {F\"oldi}},\ and\ \bibinfo {author}
  {\bibfnamefont {S.}~\bibnamefont {Varr\'o}},\ }\bibfield  {title} {\bibinfo
  {title} {Quantum-optical description of photon statistics and cross
  correlations in high-order harmonic generation},\ }\href
  {https://doi.org/10.1103/PhysRevA.104.033703} {\bibfield  {journal} {\bibinfo
   {journal} {Phys. Rev. A}\ }\textbf {\bibinfo {volume} {104}},\ \bibinfo
  {pages} {033703} (\bibinfo {year} {2021})}\BibitemShut {NoStop}%
\bibitem [{\citenamefont {Gorlach}\ \emph {et~al.}(2020)\citenamefont
  {Gorlach}, \citenamefont {Neufeld}, \citenamefont {Rivera}, \citenamefont
  {Cohen},\ and\ \citenamefont {Kaminer}}]{Gorlach2020}%
  \BibitemOpen
  \bibfield  {author} {\bibinfo {author} {\bibfnamefont {A.}~\bibnamefont
  {Gorlach}}, \bibinfo {author} {\bibfnamefont {O.}~\bibnamefont {Neufeld}},
  \bibinfo {author} {\bibfnamefont {N.}~\bibnamefont {Rivera}}, \bibinfo
  {author} {\bibfnamefont {O.}~\bibnamefont {Cohen}},\ and\ \bibinfo {author}
  {\bibfnamefont {I.}~\bibnamefont {Kaminer}},\ }\bibfield  {title} {\bibinfo
  {title} {The quantum-optical nature of high harmonic generation},\ }\href
  {https://doi.org/10.1038/s41467-020-18218-w} {\bibfield  {journal} {\bibinfo
  {journal} {Nature Communications}\ }\textbf {\bibinfo {volume} {11}},\
  \bibinfo {pages} {4598} (\bibinfo {year} {2020})}\BibitemShut {NoStop}%
\bibitem [{\citenamefont {Fang}\ \emph {et~al.}(2023)\citenamefont {Fang},
  \citenamefont {Sun}, \citenamefont {He},\ and\ \citenamefont
  {Liu}}]{PhysRevLett.130.253201}%
  \BibitemOpen
  \bibfield  {author} {\bibinfo {author} {\bibfnamefont {Y.}~\bibnamefont
  {Fang}}, \bibinfo {author} {\bibfnamefont {F.-X.}\ \bibnamefont {Sun}},
  \bibinfo {author} {\bibfnamefont {Q.}~\bibnamefont {He}},\ and\ \bibinfo
  {author} {\bibfnamefont {Y.}~\bibnamefont {Liu}},\ }\bibfield  {title}
  {\bibinfo {title} {Strong-field ionization of hydrogen atoms with quantum
  light},\ }\href {https://doi.org/10.1103/PhysRevLett.130.253201} {\bibfield
  {journal} {\bibinfo  {journal} {Phys. Rev. Lett.}\ }\textbf {\bibinfo
  {volume} {130}},\ \bibinfo {pages} {253201} (\bibinfo {year}
  {2023})}\BibitemShut {NoStop}%
\bibitem [{\citenamefont {Rivera-Dean}\ \emph {et~al.}(2022)\citenamefont
  {Rivera-Dean}, \citenamefont {Stammer}, \citenamefont {Maxwell},
  \citenamefont {Lamprou}, \citenamefont {Tzallas}, \citenamefont
  {Lewenstein},\ and\ \citenamefont {Ciappina}}]{PhysRevA.106.063705}%
  \BibitemOpen
  \bibfield  {author} {\bibinfo {author} {\bibfnamefont {J.}~\bibnamefont
  {Rivera-Dean}}, \bibinfo {author} {\bibfnamefont {P.}~\bibnamefont
  {Stammer}}, \bibinfo {author} {\bibfnamefont {A.~S.}\ \bibnamefont
  {Maxwell}}, \bibinfo {author} {\bibfnamefont {T.}~\bibnamefont {Lamprou}},
  \bibinfo {author} {\bibfnamefont {P.}~\bibnamefont {Tzallas}}, \bibinfo
  {author} {\bibfnamefont {M.}~\bibnamefont {Lewenstein}},\ and\ \bibinfo
  {author} {\bibfnamefont {M.~F.}\ \bibnamefont {Ciappina}},\ }\bibfield
  {title} {\bibinfo {title} {Light-matter entanglement after above-threshold
  ionization processes in atoms},\ }\href
  {https://doi.org/10.1103/PhysRevA.106.063705} {\bibfield  {journal} {\bibinfo
   {journal} {Phys. Rev. A}\ }\textbf {\bibinfo {volume} {106}},\ \bibinfo
  {pages} {063705} (\bibinfo {year} {2022})}\BibitemShut {NoStop}%
\bibitem [{\citenamefont {Jaynes}\ and\ \citenamefont
  {Cummings}(1962)}]{Jaynes1962ComparisonOQ}%
  \BibitemOpen
  \bibfield  {author} {\bibinfo {author} {\bibfnamefont {E.}~\bibnamefont
  {Jaynes}}\ and\ \bibinfo {author} {\bibfnamefont {F.~W.}\ \bibnamefont
  {Cummings}},\ }\bibfield  {title} {\bibinfo {title} {Comparison of quantum
  and semiclassical radiation theories with application to the beam maser}\
  }(\bibinfo {year} {1962})\BibitemShut {NoStop}%
\bibitem [{\citenamefont {Buchholz}\ \emph {et~al.}(2019)\citenamefont
  {Buchholz}, \citenamefont {Theophilou}, \citenamefont {Nielsen},
  \citenamefont {Ruggenthaler},\ and\ \citenamefont
  {Rubio}}]{doi:10.1021/acsphotonics.9b00648}%
  \BibitemOpen
  \bibfield  {author} {\bibinfo {author} {\bibfnamefont {F.}~\bibnamefont
  {Buchholz}}, \bibinfo {author} {\bibfnamefont {I.}~\bibnamefont
  {Theophilou}}, \bibinfo {author} {\bibfnamefont {S.~E.~B.}\ \bibnamefont
  {Nielsen}}, \bibinfo {author} {\bibfnamefont {M.}~\bibnamefont
  {Ruggenthaler}},\ and\ \bibinfo {author} {\bibfnamefont {A.}~\bibnamefont
  {Rubio}},\ }\bibfield  {title} {\bibinfo {title} {Reduced density-matrix
  approach to strong matter-photon interaction},\ }\href
  {https://doi.org/10.1021/acsphotonics.9b00648} {\bibfield  {journal}
  {\bibinfo  {journal} {ACS Photonics}\ }\textbf {\bibinfo {volume} {6}},\
  \bibinfo {pages} {2694} (\bibinfo {year} {2019})}\BibitemShut {NoStop}%
\bibitem [{\citenamefont {Schäfer}\ \emph {et~al.}(2019)\citenamefont
  {Schäfer}, \citenamefont {Ruggenthaler}, \citenamefont {Appel},\ and\
  \citenamefont {Rubio}}]{Schafer4883}%
  \BibitemOpen
  \bibfield  {author} {\bibinfo {author} {\bibfnamefont {C.}~\bibnamefont
  {Schäfer}}, \bibinfo {author} {\bibfnamefont {M.}~\bibnamefont
  {Ruggenthaler}}, \bibinfo {author} {\bibfnamefont {H.}~\bibnamefont
  {Appel}},\ and\ \bibinfo {author} {\bibfnamefont {A.}~\bibnamefont {Rubio}},\
  }\bibfield  {title} {\bibinfo {title} {Modification of excitation and charge
  transfer in cavity quantum-electrodynamical chemistry},\ }\href
  {https://doi.org/10.1073/pnas.1814178116} {\bibfield  {journal} {\bibinfo
  {journal} {Proceedings of the National Academy of Sciences}\ }\textbf
  {\bibinfo {volume} {116}},\ \bibinfo {pages} {4883} (\bibinfo {year}
  {2019})}\BibitemShut {NoStop}%
\bibitem [{\citenamefont {Flick}\ \emph {et~al.}(2015)\citenamefont {Flick},
  \citenamefont {Ruggenthaler}, \citenamefont {Appel},\ and\ \citenamefont
  {Rubio}}]{Flick15285}%
  \BibitemOpen
  \bibfield  {author} {\bibinfo {author} {\bibfnamefont {J.}~\bibnamefont
  {Flick}}, \bibinfo {author} {\bibfnamefont {M.}~\bibnamefont {Ruggenthaler}},
  \bibinfo {author} {\bibfnamefont {H.}~\bibnamefont {Appel}},\ and\ \bibinfo
  {author} {\bibfnamefont {A.}~\bibnamefont {Rubio}},\ }\bibfield  {title}
  {\bibinfo {title} {Kohn–sham approach to quantum electrodynamical
  density-functional theory: Exact time-dependent effective potentials in real
  space},\ }\href {https://doi.org/10.1073/pnas.1518224112} {\bibfield
  {journal} {\bibinfo  {journal} {Proceedings of the National Academy of
  Sciences}\ }\textbf {\bibinfo {volume} {112}},\ \bibinfo {pages} {15285}
  (\bibinfo {year} {2015})}\BibitemShut {NoStop}%
\bibitem [{\citenamefont {Flick}\ \emph {et~al.}(2017)\citenamefont {Flick},
  \citenamefont {Ruggenthaler}, \citenamefont {Appel},\ and\ \citenamefont
  {Rubio}}]{Flick3026}%
  \BibitemOpen
  \bibfield  {author} {\bibinfo {author} {\bibfnamefont {J.}~\bibnamefont
  {Flick}}, \bibinfo {author} {\bibfnamefont {M.}~\bibnamefont {Ruggenthaler}},
  \bibinfo {author} {\bibfnamefont {H.}~\bibnamefont {Appel}},\ and\ \bibinfo
  {author} {\bibfnamefont {A.}~\bibnamefont {Rubio}},\ }\bibfield  {title}
  {\bibinfo {title} {Atoms and molecules in cavities, from weak to strong
  coupling in quantum-electrodynamics (qed) chemistry},\ }\href
  {https://doi.org/10.1073/pnas.1615509114} {\bibfield  {journal} {\bibinfo
  {journal} {Proceedings of the National Academy of Sciences}\ }\textbf
  {\bibinfo {volume} {114}},\ \bibinfo {pages} {3026} (\bibinfo {year}
  {2017})}\BibitemShut {NoStop}%
\bibitem [{\citenamefont {Rokaj}\ \emph {et~al.}(2018)\citenamefont {Rokaj},
  \citenamefont {Welakuh}, \citenamefont {Ruggenthaler},\ and\ \citenamefont
  {Rubio}}]{Rokaj_2018}%
  \BibitemOpen
  \bibfield  {author} {\bibinfo {author} {\bibfnamefont {V.}~\bibnamefont
  {Rokaj}}, \bibinfo {author} {\bibfnamefont {D.~M.}\ \bibnamefont {Welakuh}},
  \bibinfo {author} {\bibfnamefont {M.}~\bibnamefont {Ruggenthaler}},\ and\
  \bibinfo {author} {\bibfnamefont {A.}~\bibnamefont {Rubio}},\ }\bibfield
  {title} {\bibinfo {title} {Light{\textendash}matter interaction in the
  long-wavelength limit: no ground-state without dipole self-energy},\ }\href
  {https://doi.org/10.1088/1361-6455/aa9c99} {\bibfield  {journal} {\bibinfo
  {journal} {Journal of Physics B: Atomic, Molecular and Optical Physics}\
  }\textbf {\bibinfo {volume} {51}},\ \bibinfo {pages} {034005} (\bibinfo
  {year} {2018})}\BibitemShut {NoStop}%
\bibitem [{\citenamefont {Rivera}\ \emph {et~al.}(2019)\citenamefont {Rivera},
  \citenamefont {Flick},\ and\ \citenamefont
  {Narang}}]{PhysRevLett.122.193603}%
  \BibitemOpen
  \bibfield  {author} {\bibinfo {author} {\bibfnamefont {N.}~\bibnamefont
  {Rivera}}, \bibinfo {author} {\bibfnamefont {J.}~\bibnamefont {Flick}},\ and\
  \bibinfo {author} {\bibfnamefont {P.}~\bibnamefont {Narang}},\ }\bibfield
  {title} {\bibinfo {title} {Variational theory of nonrelativistic quantum
  electrodynamics},\ }\href {https://doi.org/10.1103/PhysRevLett.122.193603}
  {\bibfield  {journal} {\bibinfo  {journal} {Phys. Rev. Lett.}\ }\textbf
  {\bibinfo {volume} {122}},\ \bibinfo {pages} {193603} (\bibinfo {year}
  {2019})}\BibitemShut {NoStop}%
\bibitem [{\citenamefont {Flick}\ and\ \citenamefont
  {Narang}(2018)}]{PhysRevLett.121.113002}%
  \BibitemOpen
  \bibfield  {author} {\bibinfo {author} {\bibfnamefont {J.}~\bibnamefont
  {Flick}}\ and\ \bibinfo {author} {\bibfnamefont {P.}~\bibnamefont {Narang}},\
  }\bibfield  {title} {\bibinfo {title} {Cavity-correlated electron-nuclear
  dynamics from first principles},\ }\href
  {https://doi.org/10.1103/PhysRevLett.121.113002} {\bibfield  {journal}
  {\bibinfo  {journal} {Phys. Rev. Lett.}\ }\textbf {\bibinfo {volume} {121}},\
  \bibinfo {pages} {113002} (\bibinfo {year} {2018})}\BibitemShut {NoStop}%
\bibitem [{\citenamefont {Hoffmann}\ \emph {et~al.}(2020)\citenamefont
  {Hoffmann}, \citenamefont {Lacombe}, \citenamefont {Rubio},\ and\
  \citenamefont {Maitra}}]{doi:10.1063/5.0012723}%
  \BibitemOpen
  \bibfield  {author} {\bibinfo {author} {\bibfnamefont {N.~M.}\ \bibnamefont
  {Hoffmann}}, \bibinfo {author} {\bibfnamefont {L.}~\bibnamefont {Lacombe}},
  \bibinfo {author} {\bibfnamefont {A.}~\bibnamefont {Rubio}},\ and\ \bibinfo
  {author} {\bibfnamefont {N.~T.}\ \bibnamefont {Maitra}},\ }\bibfield  {title}
  {\bibinfo {title} {Effect of many modes on self-polarization and
  photochemical suppression in cavities},\ }\href
  {https://doi.org/10.1063/5.0012723} {\bibfield  {journal} {\bibinfo
  {journal} {The Journal of Chemical Physics}\ }\textbf {\bibinfo {volume}
  {153}},\ \bibinfo {pages} {104103} (\bibinfo {year} {2020})}\BibitemShut
  {NoStop}%
\bibitem [{\citenamefont {Tokatly}(2018)}]{PhysRevB.98.235123}%
  \BibitemOpen
  \bibfield  {author} {\bibinfo {author} {\bibfnamefont {I.~V.}\ \bibnamefont
  {Tokatly}},\ }\bibfield  {title} {\bibinfo {title} {Conserving approximations
  in cavity quantum electrodynamics: Implications for density functional theory
  of electron-photon systems},\ }\href
  {https://doi.org/10.1103/PhysRevB.98.235123} {\bibfield  {journal} {\bibinfo
  {journal} {Phys. Rev. B}\ }\textbf {\bibinfo {volume} {98}},\ \bibinfo
  {pages} {235123} (\bibinfo {year} {2018})}\BibitemShut {NoStop}%
\bibitem [{\citenamefont {Galego}\ \emph {et~al.}(2017)\citenamefont {Galego},
  \citenamefont {Garcia-Vidal},\ and\ \citenamefont
  {Feist}}]{PhysRevLett.119.136001}%
  \BibitemOpen
  \bibfield  {author} {\bibinfo {author} {\bibfnamefont {J.}~\bibnamefont
  {Galego}}, \bibinfo {author} {\bibfnamefont {F.~J.}\ \bibnamefont
  {Garcia-Vidal}},\ and\ \bibinfo {author} {\bibfnamefont {J.}~\bibnamefont
  {Feist}},\ }\bibfield  {title} {\bibinfo {title} {Many-molecule reaction
  triggered by a single photon in polaritonic chemistry},\ }\href
  {https://doi.org/10.1103/PhysRevLett.119.136001} {\bibfield  {journal}
  {\bibinfo  {journal} {Phys. Rev. Lett.}\ }\textbf {\bibinfo {volume} {119}},\
  \bibinfo {pages} {136001} (\bibinfo {year} {2017})}\BibitemShut {NoStop}%
\bibitem [{\citenamefont {Mandal}\ \emph
  {et~al.}(2020{\natexlab{a}})\citenamefont {Mandal}, \citenamefont
  {Montillo~Vega},\ and\ \citenamefont {Huo}}]{Mandal}%
  \BibitemOpen
  \bibfield  {author} {\bibinfo {author} {\bibfnamefont {A.}~\bibnamefont
  {Mandal}}, \bibinfo {author} {\bibfnamefont {S.}~\bibnamefont
  {Montillo~Vega}},\ and\ \bibinfo {author} {\bibfnamefont {P.}~\bibnamefont
  {Huo}},\ }\bibfield  {title} {\bibinfo {title} {Polarized fock states and the
  dynamical casimir effect in molecular cavity quantum electrodynamics},\
  }\href {https://doi.org/10.1021/acs.jpclett.0c02399} {\bibfield  {journal}
  {\bibinfo  {journal} {The Journal of Physical Chemistry Letters}\ }\textbf
  {\bibinfo {volume} {11}},\ \bibinfo {pages} {9215} (\bibinfo {year}
  {2020}{\natexlab{a}})},\ \bibinfo {note} {pMID: 32991814}\BibitemShut
  {NoStop}%
\bibitem [{\citenamefont {Cederbaum}\ and\ \citenamefont
  {Kuleff}(2021)}]{Cederbaum2021}%
  \BibitemOpen
  \bibfield  {author} {\bibinfo {author} {\bibfnamefont {L.~S.}\ \bibnamefont
  {Cederbaum}}\ and\ \bibinfo {author} {\bibfnamefont {A.~I.}\ \bibnamefont
  {Kuleff}},\ }\bibfield  {title} {\bibinfo {title} {Impact of cavity on
  interatomic coulombic decay},\ }\href
  {https://doi.org/10.1038/s41467-021-24221-6} {\bibfield  {journal} {\bibinfo
  {journal} {Nature Communications}\ }\textbf {\bibinfo {volume} {12}},\
  \bibinfo {pages} {4083} (\bibinfo {year} {2021})}\BibitemShut {NoStop}%
\bibitem [{\citenamefont {Szidarovszky}\ \emph {et~al.}(2018)\citenamefont
  {Szidarovszky}, \citenamefont {Hal\'asz}, \citenamefont {Cs\'asz\'ar},
  \citenamefont {Cederbaum},\ and\ \citenamefont
  {Vib\'ok}}]{doi:10.1021/acs.jpclett.8b02609}%
  \BibitemOpen
  \bibfield  {author} {\bibinfo {author} {\bibfnamefont {T.}~\bibnamefont
  {Szidarovszky}}, \bibinfo {author} {\bibfnamefont {G.~J.}\ \bibnamefont
  {Hal\'asz}}, \bibinfo {author} {\bibfnamefont {A.~G.}\ \bibnamefont
  {Cs\'asz\'ar}}, \bibinfo {author} {\bibfnamefont {L.~S.}\ \bibnamefont
  {Cederbaum}},\ and\ \bibinfo {author} {\bibfnamefont {A.}~\bibnamefont
  {Vib\'ok}},\ }\bibfield  {title} {\bibinfo {title} {Conical intersections
  induced by quantum light: Field-dressed spectra from the weak to the
  ultrastrong coupling regimes},\ }\href
  {https://doi.org/10.1021/acs.jpclett.8b02609} {\bibfield  {journal} {\bibinfo
   {journal} {The Journal of Physical Chemistry Letters}\ }\textbf {\bibinfo
  {volume} {9}},\ \bibinfo {pages} {6215} (\bibinfo {year} {2018})}\BibitemShut
  {NoStop}%
\bibitem [{\citenamefont {Ashida}\ and\ \citenamefont
  {Demler}(2021)}]{PhysRevLett.126.153603}%
  \BibitemOpen
  \bibfield  {author} {\bibinfo {author} {\bibfnamefont {A.~I.}\ \bibnamefont
  {Ashida}, \bibfnamefont {Yuto}}\ and\ \bibinfo {author} {\bibfnamefont
  {E.}~\bibnamefont {Demler}},\ }\bibfield  {title} {\bibinfo {title} {Cavity
  quantum electrodynamics at arbitrary light-matter coupling strengths},\
  }\href {https://doi.org/10.1103/PhysRevLett.126.153603} {\bibfield  {journal}
  {\bibinfo  {journal} {Phys. Rev. Lett.}\ }\textbf {\bibinfo {volume} {126}},\
  \bibinfo {pages} {153603} (\bibinfo {year} {2021})}\BibitemShut {NoStop}%
\bibitem [{\citenamefont {Craig}\ and\ \citenamefont
  {Thirunamachandran}(1984)}]{craig1984molecular}%
  \BibitemOpen
  \bibfield  {author} {\bibinfo {author} {\bibfnamefont {D.}~\bibnamefont
  {Craig}}\ and\ \bibinfo {author} {\bibfnamefont {T.}~\bibnamefont
  {Thirunamachandran}},\ }\href
  {https://books.google.com/books?id=SAHmzAEACAAJ} {\emph {\bibinfo {title}
  {Molecular Quantum Electrodynamics; An Introduction to Radiation Molecule
  Interactions}}}\ (\bibinfo {year} {1984})\BibitemShut {NoStop}%
\bibitem [{\citenamefont {Mandal}\ \emph
  {et~al.}(2020{\natexlab{b}})\citenamefont {Mandal}, \citenamefont {Krauss},\
  and\ \citenamefont {Huo}}]{acs.jpcb.0c03227}%
  \BibitemOpen
  \bibfield  {author} {\bibinfo {author} {\bibfnamefont {A.}~\bibnamefont
  {Mandal}}, \bibinfo {author} {\bibfnamefont {T.~D.}\ \bibnamefont {Krauss}},\
  and\ \bibinfo {author} {\bibfnamefont {P.}~\bibnamefont {Huo}},\ }\bibfield
  {title} {\bibinfo {title} {Polariton-mediated electron transfer via cavity
  quantum electrodynamics},\ }\href {https://doi.org/10.1021/acs.jpcb.0c03227}
  {\bibfield  {journal} {\bibinfo  {journal} {The Journal of Physical Chemistry
  B}\ }\textbf {\bibinfo {volume} {124}},\ \bibinfo {pages} {6321} (\bibinfo
  {year} {2020}{\natexlab{b}})},\ \bibinfo {note} {pMID: 32589846}\BibitemShut
  {NoStop}%
\bibitem [{\citenamefont {Taylor}\ \emph {et~al.}(2022)\citenamefont {Taylor},
  \citenamefont {Mandal},\ and\ \citenamefont {Huo}}]{Taylor:22}%
  \BibitemOpen
  \bibfield  {author} {\bibinfo {author} {\bibfnamefont {M.~A.~D.}\
  \bibnamefont {Taylor}}, \bibinfo {author} {\bibfnamefont {A.}~\bibnamefont
  {Mandal}},\ and\ \bibinfo {author} {\bibfnamefont {P.}~\bibnamefont {Huo}},\
  }\bibfield  {title} {\bibinfo {title} {Resolving ambiguities of the mode
  truncation in cavity quantum electrodynamics},\ }\href
  {https://doi.org/10.1364/OL.450228} {\bibfield  {journal} {\bibinfo
  {journal} {Opt. Lett.}\ }\textbf {\bibinfo {volume} {47}},\ \bibinfo {pages}
  {1446} (\bibinfo {year} {2022})}\BibitemShut {NoStop}%
\bibitem [{\citenamefont {Petersson}\ and\ \citenamefont
  {Hellsing}(2009)}]{Petersson_2010}%
  \BibitemOpen
  \bibfield  {author} {\bibinfo {author} {\bibfnamefont {T.}~\bibnamefont
  {Petersson}}\ and\ \bibinfo {author} {\bibfnamefont {B.}~\bibnamefont
  {Hellsing}},\ }\bibfield  {title} {\bibinfo {title} {A detailed derivation of
  gaussian orbital-based matrix elements in electron structure calculations},\
  }\href {https://doi.org/10.1088/0143-0807/31/1/004} {\bibfield  {journal}
  {\bibinfo  {journal} {European Journal of Physics}\ }\textbf {\bibinfo
  {volume} {31}},\ \bibinfo {pages} {37} (\bibinfo {year} {2009})}\BibitemShut
  {NoStop}%
\bibitem [{\citenamefont {Boys}\ and\ \citenamefont
  {Egerton}(1950)}]{doi:10.1098/rspa.1950.0036}%
  \BibitemOpen
  \bibfield  {author} {\bibinfo {author} {\bibfnamefont {S.~F.}\ \bibnamefont
  {Boys}}\ and\ \bibinfo {author} {\bibfnamefont {A.~C.}\ \bibnamefont
  {Egerton}},\ }\bibfield  {title} {\bibinfo {title} {Electronic wave functions
  - i. a general method of calculation for the stationary states of any
  molecular system},\ }\href {https://doi.org/10.1098/rspa.1950.0036}
  {\bibfield  {journal} {\bibinfo  {journal} {Proceedings of the Royal Society
  of London. Series A. Mathematical and Physical Sciences}\ }\textbf {\bibinfo
  {volume} {200}},\ \bibinfo {pages} {542} (\bibinfo {year}
  {1950})}\BibitemShut {NoStop}%
\bibitem [{\citenamefont {Mitroy}\ \emph {et~al.}(2013)\citenamefont {Mitroy},
  \citenamefont {Bubin}, \citenamefont {Horiuchi}, \citenamefont {Suzuki},
  \citenamefont {Adamowicz}, \citenamefont {Cencek}, \citenamefont {Szalewicz},
  \citenamefont {Komasa}, \citenamefont {Blume},\ and\ \citenamefont
  {Varga}}]{RevModPhys.85.693}%
  \BibitemOpen
  \bibfield  {author} {\bibinfo {author} {\bibfnamefont {J.}~\bibnamefont
  {Mitroy}}, \bibinfo {author} {\bibfnamefont {S.}~\bibnamefont {Bubin}},
  \bibinfo {author} {\bibfnamefont {W.}~\bibnamefont {Horiuchi}}, \bibinfo
  {author} {\bibfnamefont {Y.}~\bibnamefont {Suzuki}}, \bibinfo {author}
  {\bibfnamefont {L.}~\bibnamefont {Adamowicz}}, \bibinfo {author}
  {\bibfnamefont {W.}~\bibnamefont {Cencek}}, \bibinfo {author} {\bibfnamefont
  {K.}~\bibnamefont {Szalewicz}}, \bibinfo {author} {\bibfnamefont
  {J.}~\bibnamefont {Komasa}}, \bibinfo {author} {\bibfnamefont
  {D.}~\bibnamefont {Blume}},\ and\ \bibinfo {author} {\bibfnamefont
  {K.}~\bibnamefont {Varga}},\ }\bibfield  {title} {\bibinfo {title} {Theory
  and application of explicitly correlated gaussians},\ }\href
  {https://doi.org/10.1103/RevModPhys.85.693} {\bibfield  {journal} {\bibinfo
  {journal} {Rev. Mod. Phys.}\ }\textbf {\bibinfo {volume} {85}},\ \bibinfo
  {pages} {693} (\bibinfo {year} {2013})}\BibitemShut {NoStop}%
\bibitem [{\citenamefont {Suzuki}\ \emph {et~al.}(1998)\citenamefont {Suzuki},
  ,\ and\ \citenamefont {Varga}}]{suzuki1998stochastic}%
  \BibitemOpen
  \bibfield  {author} {\bibinfo {author} {\bibfnamefont {Y.}~\bibnamefont
  {Suzuki}}, ,\ and\ \bibinfo {author} {\bibfnamefont {K.}~\bibnamefont
  {Varga}},\ }\href@noop {} {\emph {\bibinfo {title} {Stochastic variational
  approach to quantum-mechanical few-body problems}}},\ Vol.~\bibinfo {volume}
  {54}\ (\bibinfo  {publisher} {Springer Science \& Business Media},\ \bibinfo
  {year} {1998})\BibitemShut {NoStop}%
\bibitem [{\citenamefont {Heiss}\ and\ \citenamefont
  {Sannino}(1990)}]{Heiss_1990}%
  \BibitemOpen
  \bibfield  {author} {\bibinfo {author} {\bibfnamefont {W.~D.}\ \bibnamefont
  {Heiss}}\ and\ \bibinfo {author} {\bibfnamefont {A.~L.}\ \bibnamefont
  {Sannino}},\ }\bibfield  {title} {\bibinfo {title} {Avoided level crossing
  and exceptional points},\ }\href {https://doi.org/10.1088/0305-4470/23/7/022}
  {\bibfield  {journal} {\bibinfo  {journal} {Journal of Physics A:
  Mathematical and General}\ }\textbf {\bibinfo {volume} {23}},\ \bibinfo
  {pages} {1167} (\bibinfo {year} {1990})}\BibitemShut {NoStop}%
\bibitem [{\citenamefont {Eleuch}\ and\ \citenamefont
  {Rotter}(2013)}]{https://doi.org/10.1002/prop.201200062}%
  \BibitemOpen
  \bibfield  {author} {\bibinfo {author} {\bibfnamefont {H.}~\bibnamefont
  {Eleuch}}\ and\ \bibinfo {author} {\bibfnamefont {I.}~\bibnamefont
  {Rotter}},\ }\bibfield  {title} {\bibinfo {title} {Avoided level crossings in
  open quantum systems},\ }\href
  {https://doi.org/https://doi.org/10.1002/prop.201200062} {\bibfield
  {journal} {\bibinfo  {journal} {Fortschritte der Physik}\ }\textbf {\bibinfo
  {volume} {61}},\ \bibinfo {pages} {194} (\bibinfo {year} {2013})}\BibitemShut
  {NoStop}%
\bibitem [{\citenamefont {Rotter}(2001)}]{PhysRevE.64.036213}%
  \BibitemOpen
  \bibfield  {author} {\bibinfo {author} {\bibfnamefont {I.}~\bibnamefont
  {Rotter}},\ }\bibfield  {title} {\bibinfo {title} {Dynamics of quantum
  systems},\ }\href {https://doi.org/10.1103/PhysRevE.64.036213} {\bibfield
  {journal} {\bibinfo  {journal} {Phys. Rev. E}\ }\textbf {\bibinfo {volume}
  {64}},\ \bibinfo {pages} {036213} (\bibinfo {year} {2001})}\BibitemShut
  {NoStop}%
\bibitem [{\citenamefont {Krause}\ \emph {et~al.}(1992)\citenamefont {Krause},
  \citenamefont {Schafer},\ and\ \citenamefont {Kulander}}]{PhysRevA.45.4998}%
  \BibitemOpen
  \bibfield  {author} {\bibinfo {author} {\bibfnamefont {J.~L.}\ \bibnamefont
  {Krause}}, \bibinfo {author} {\bibfnamefont {K.~J.}\ \bibnamefont
  {Schafer}},\ and\ \bibinfo {author} {\bibfnamefont {K.~C.}\ \bibnamefont
  {Kulander}},\ }\bibfield  {title} {\bibinfo {title} {Calculation of
  photoemission from atoms subject to intense laser fields},\ }\href
  {https://doi.org/10.1103/PhysRevA.45.4998} {\bibfield  {journal} {\bibinfo
  {journal} {Phys. Rev. A}\ }\textbf {\bibinfo {volume} {45}},\ \bibinfo
  {pages} {4998} (\bibinfo {year} {1992})}\BibitemShut {NoStop}%
\bibitem [{\citenamefont {Lewenstein}\ \emph {et~al.}(1994)\citenamefont
  {Lewenstein}, \citenamefont {Balcou}, \citenamefont {Ivanov}, \citenamefont
  {L'Huillier},\ and\ \citenamefont {Corkum}}]{PhysRevA.49.2117}%
  \BibitemOpen
  \bibfield  {author} {\bibinfo {author} {\bibfnamefont {M.}~\bibnamefont
  {Lewenstein}}, \bibinfo {author} {\bibfnamefont {P.}~\bibnamefont {Balcou}},
  \bibinfo {author} {\bibfnamefont {M.~Y.}\ \bibnamefont {Ivanov}}, \bibinfo
  {author} {\bibfnamefont {A.}~\bibnamefont {L'Huillier}},\ and\ \bibinfo
  {author} {\bibfnamefont {P.~B.}\ \bibnamefont {Corkum}},\ }\bibfield  {title}
  {\bibinfo {title} {Theory of high-harmonic generation by low-frequency laser
  fields},\ }\href {https://doi.org/10.1103/PhysRevA.49.2117} {\bibfield
  {journal} {\bibinfo  {journal} {Phys. Rev. A}\ }\textbf {\bibinfo {volume}
  {49}},\ \bibinfo {pages} {2117} (\bibinfo {year} {1994})}\BibitemShut
  {NoStop}%
\end{thebibliography}
%

\end{document}